\documentclass[twocolumn,prl,floatfix,superscriptaddress,longbibliography]{revtex4-2}

\usepackage[hidelinks,colorlinks=true,citecolor=blue]{hyperref}
\usepackage{float}
\usepackage{upgreek}
\usepackage{graphicx} 
\usepackage{gensymb}
\usepackage{epsfig}
\usepackage{amsmath,bm,amssymb}
\usepackage{amsfonts}
\usepackage[usenames]{color}
\usepackage{times}
\usepackage{bm}
\usepackage{xspace}

\newcommand{\Tc}{\ensuremath{T_{\mathrm{c}}}\xspace}
\newcommand{\asp}{\ensuremath{\alpha_{\mathrm{sp}}}\xspace}

\begin{document}

\title{Quenched pair breaking by interlayer correlations as a key to superconductivity in La$_3$Ni$_2$O$_7$}

 \author{Siheon Ryee}
 \email{sryee@physnet.uni-hamburg.de} 
 \affiliation{I. Institute of Theoretical Physics, University of Hamburg, Notkestrasse 9, 22607 Hamburg, Germany}

 \author{Niklas Witt}
 \affiliation{I. Institute of Theoretical Physics, University of Hamburg, Notkestrasse 9, 22607 Hamburg, Germany}
 \affiliation{The Hamburg Centre for Ultrafast Imaging, Luruper Chaussee 149, 22761 Hamburg, Germany}

 \author{Tim O. Wehling}
 \affiliation{I. Institute of Theoretical Physics, University of Hamburg, Notkestrasse 9, 22607 Hamburg, Germany}
 \affiliation{The Hamburg Centre for Ultrafast Imaging, Luruper Chaussee 149, 22761 Hamburg, Germany}
 
\date{\today}

\begin{abstract}
The recent discovery of superconductivity in La$_3$Ni$_2$O$_7$ with $T_\mathrm{c} \simeq 80$\,K under high pressure opens up a new route to high-$T_\mathrm{c}$ superconductivity. This material realizes a bilayer square lattice model featuring a strong interlayer hybridization unlike many unconventional superconductors. A key question in this regard concerns how electronic correlations driven by the interlayer hybridization affect the low-energy electronic structure and the concomitant superconductivity. Here, we demonstrate using a cluster dynamical mean-field theory that the interlayer electronic correlations (IECs) induce a Lifshitz transition resulting in a change of Fermi surface topology. By solving an appropriate gap equation, we further show that the leading pairing instability, $s \pm$-wave, is enhanced by the IECs. The underlying mechanism is the quenching of a strong ferromagnetic channel, resulting from the Lifshitz transition driven by the IECs. Based on this picture, we provide a possible reason of why superconductivity emerges only under high pressure. 
\end{abstract}
 
\maketitle

\begin{figure} [!htbp] 
	\includegraphics[width=0.9\columnwidth, angle=0]{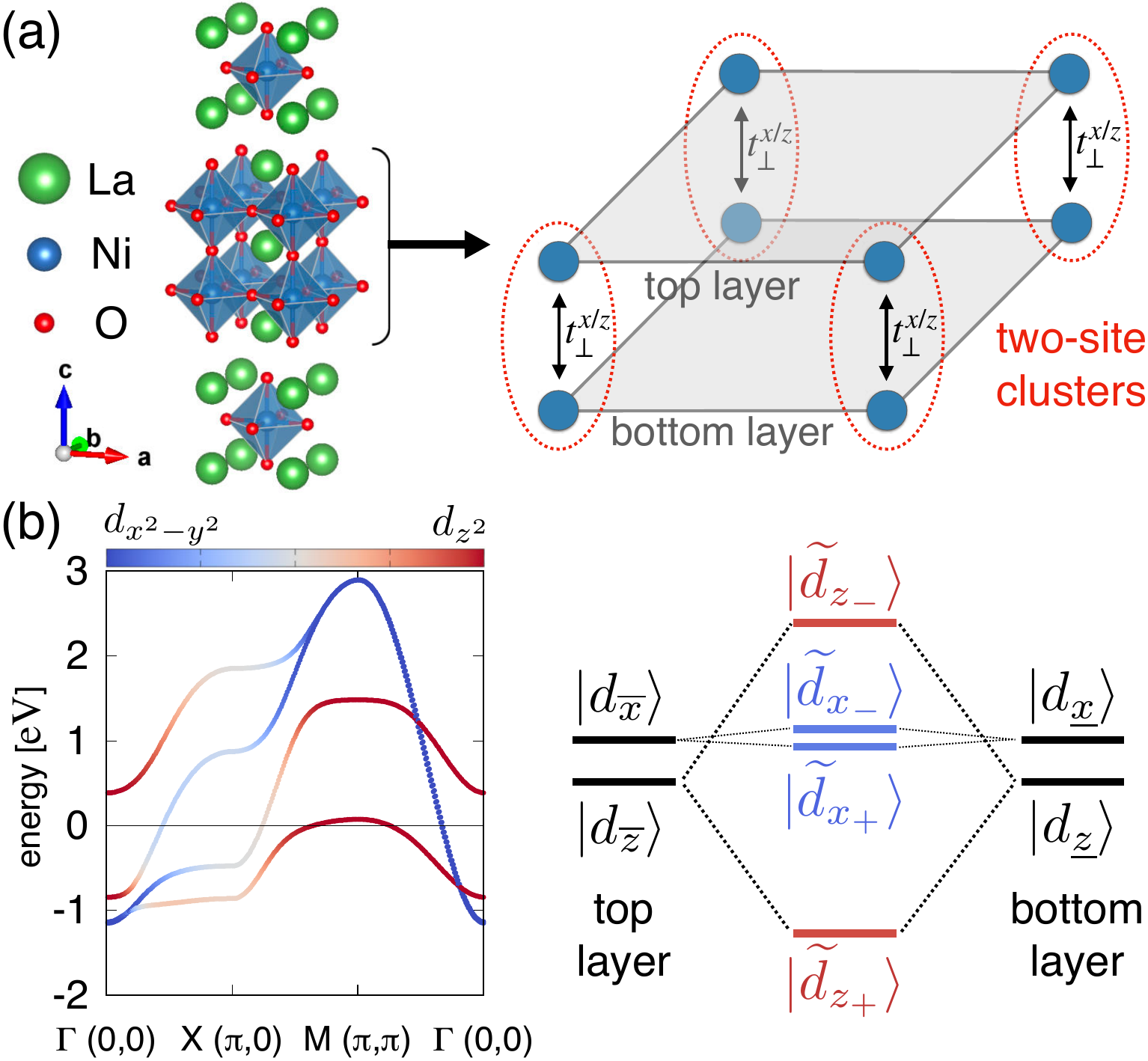}
	\caption{(a) Left: Crystal structure of La$_3$Ni$_2$O$_7$ under high pressure drawn using VESTA \cite{VESTA}. Right: The bilayer square lattice model for La$_3$Ni$_2$O$_7$.  Dimers consisting of top and bottom layer Ni sites (blue circles) coupled via $t^{x/z}_{\perp}$ are highlighted with red-dotted ovals. (b) Left: MLWF bands of the DFT electronic structure. The color bar indicates the orbital character. Right: Sketch of the formation of four BA orbitals within the dimer consisting of the Ni-$e_g$ orbitals in the top and bottom layers. 
	}
	\label{fig1}
\end{figure}

The recent discovery of superconductivity in bilayer nickelate La$_3$Ni$_2$O$_7$ under high pressure [Fig.~\ref{fig1}(a)] heralds a new class of high-\Tc superconductors~\cite{sun_signatures_2023}. Without doping, this material exhibits superconductivity under pressure exceeding \mbox{$14$\,GPa} with maximal critical temperature of \mbox{$\Tc \simeq 80$\,K}~\cite{sun_signatures_2023,hou_emergence_2023,zhang_high-temperature_2023,wang_pressure-induced_2023,zhang_effects_2023}. 
A notable feature in La$_3$Ni$_2$O$_7$ is a multiorbital nature of low-lying states already at the level of density functional theory (DFT)~\cite{sun_signatures_2023,luo_bilayer_2023,shilenko_correlated_2023,gu_effective_2023,lechermann_electronic_2023,zhang_electronic_2023,cao_flat_2023,yang_possible_2023,sakakibara_possible_2023,jiang_pressure_2023,rhodes_structural_2023,zhang_electronic_2023,zhang_trends_2023,geisler_structural_2023,liu_spm-wave_2023,labollita_electronic_2023,sakakibara_theoretical_2023} [Fig.~\ref{fig1}(b)].  Namely, 
three electrons per unit cell are distributed over Ni-$e_g$ orbitals in the top and bottom square-planar lattices, whereas Ni-$t_{2g}$ orbitals are fully occupied, thereby inactive for the low-energy physics. The two layers are coupled dominantly via interlayer nearest-neighbor hopping (or hybridization) between Ni-$d_{z^2}$ orbitals ($t_{\perp}^{z} \simeq -0.63$\,eV)~\cite{luo_bilayer_2023,gu_effective_2023,sakakibara_possible_2023}. The hopping between Ni-$d_{x^2-y^2}$ is much smaller ($|t_{\perp}^{x}| < 0.05$\,eV)~\cite{luo_bilayer_2023,gu_effective_2023}. Most importantly, $t_{\perp}^{z}$ is deemed to be crucial for the noninteracting Fermi surface (FS) topology and theories of superconductivity in La$_3$Ni$_2$O$_7$~\cite{liu_spm-wave_2023,zhang_structural_2023,zhang_trends_2023,sakakibara_possible_2023,oh_type_2023,liao_electron_2023,qin_high-t_c_2023,lu_interlayer_2023,tian_correlation_2023,yang_possible_2023,lu_superconductivity_2023,lange_feshbach_2023,wu_charge_2023,labollita_electronic_2023}.

In this respect, an important open question concerns how interlayer electronic correlations (IECs) driven by $t^{x/z}_{\perp}$ modify the low-energy electronic structure and how they affect superconductivity. 
Since $t^{z}_{\perp}$ is the largest among all the hopping amplitudes \cite{luo_bilayer_2023,gu_effective_2023}, one can identify the interlayer nearest-neighbor electronic correlations in the Ni-$d_{z^2}$ states as the leading ``nonlocal" correlations.

In this Letter, we employ a cluster (cellular) dynamical mean-field theory (CDMFT) \cite{Lichtenstein2000,CDMFT,Maier2005} to address nonperturbatively the nonlocal as well as the local electronic correlations within the two-site clusters (dimers) of the bilayer square lattice model for La$_3$Ni$_2$O$_7$ [Fig.~\ref{fig1}(a)]. One of the key findings of our study is a Lifshitz transition resulting in a change of the FS topology which does not occur when only local correlations are taken into account. By solving an appropriate gap equation, we show that the IECs promote $s\pm$-wave pairing. The underlying mechanism is the quenching of ferromagnetic (FM) fluctuations resulting from the Lifshitz transition due to IECs. Based on this picture, we provide a possible reason of why superconductivity emerges only under high pressure.

We consider a Hamiltonian on the bilayer square lattice: $\mathcal{H}=H_0 + H_\mathrm{int}$. Here, $H_0$ is a tight-binding term for the Ni-$e_g$ subspace describing the band structure for which we use the maximally localized Wannier function (MLWF) description for the DFT result of La$_3$Ni$_2$O$_7$ under high pressure (29.5~GPa) \cite{gu_effective_2023} [Fig.~\ref{fig1}(b)].  $H_\mathrm{int}$ is the local interaction term between Ni-$e_g$ orbital electrons on the same Ni site, and is given by the standard Kanamori form consisting of $U$ (intraorbital Coulomb interaction),  $J$ (Hund's coupling), and $U'$ (interorbital Coulomb interaction; $U'=U-2J$). We use $U=3.7$, $J =0.6$, and $U'=2.5$~eV by taking {\it ab initio} estimates for the $e_g$ MLWF model \cite{christiansson_correlated_2023}. 
See Supplemental Material for more information \cite{supple}.

The  impurity problem is solved using the hybridization-expansion continuous-time quantum Monte Carlo method \cite{CTQMC,comctqmc}. We investigate the system at a temperature of $T=1/145~\mathrm{eV} \simeq 80$~K corresponding to the maximum experimental \Tc \cite{sun_signatures_2023}. To mitigate the Monte Carlo sign problem resulting from the large interlayer hybridization in CDMFT, we solve the model in a bonding-antibonding (BA) basis defined as the $+$ or $-$ combinations of the top and bottom layer $e_g$ orbitals:
\begin{align} 
|\widetilde{d}_{i\eta_\pm \sigma} \rangle= ( |d_{i\bar{\eta} \sigma} \rangle \pm  | d_{i {\underline{\eta}} \sigma} \rangle )/\sqrt{2}\;.
\end{align}
Here, ket symbols indicate the corresponding Wannier states with site index $i$ for the bilayer square lattice and spin $\sigma \in \{ \uparrow, \downarrow \}$. $\bar{\eta}$ and $\underline{\eta}$ represent Ni-$e_g$ orbitals ($\eta \in  \{ x^2-y^2, z^2 \}$) in the top and bottom layers, respectively.   Hereafter $x^2-y^2$ is denoted by $x$ and $z^2$ by $z$. Site and spin indices are omitted unless needed.
In this BA basis the CDMFT self-energy becomes orbital diagonal and momentum independent, so our CDMFT is equivalent to ``four-orbital single-site DMFT". The interlayer hopping $t^\eta_{\perp}$ between $e_g$ orbitals $|d_{\bar{\eta}} \rangle $ and $|d_{\underline{\eta}} \rangle$ turns into a hybridization gap of $2|t^\eta_{\perp}|$  between BA orbitals $|\widetilde{d}_{\eta_+} \rangle$ and $|\widetilde{d}_{\eta_-}  \rangle$. Thus a small (large) splitting is realized for $\eta = x$ ($\eta = z$) [schematically shown in the right panel of Fig.~\ref{fig1}(b)]. 

We first investigate how interlayer correlations affect the low-energy electronic structure by contrasting DMFT (in which all the interlayer correlations are neglected \cite{DMFT}) and CDMFT results for the same model. Note that, in a reasonable range around the {\it ab initio} interaction parameters, neither a Mott transition nor a bad metal behavior emerges within our calculations \cite{supple}, which is in line with experiments \cite{sun_signatures_2023,hou_emergence_2023,zhang_high-temperature_2023,wang_pressure-induced_2023}.

\begin{figure} [!htbp]	 
	\includegraphics[width=0.98\columnwidth]{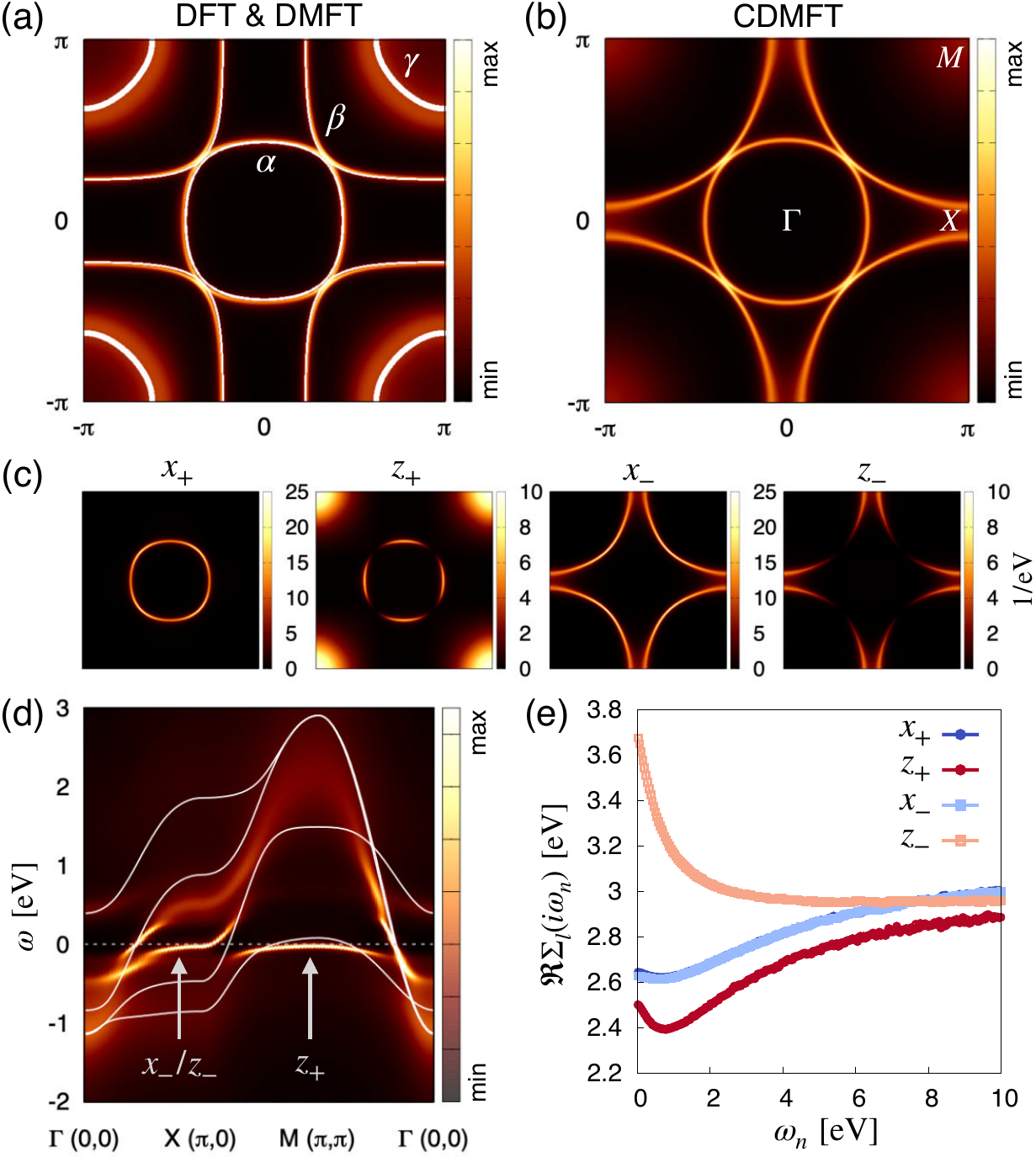}
	\caption{(a) FSs obtained from the MLWF model of the DFT band structure (white lines) and DMFT (color map). (b) FS from CDMFT. The FSs of DMFT and CDMFT in (a) and (b) are approximated via $-\sum_{lm}\delta_{l m}\Im G_{l m}(\bm{k},i\omega_0)$ where $l,m \in \{x_+,z_+,x_-,z_- \}$. (c) The orbital character of the CDMFT FS. (d) The momentum-dependent spectral function obtained from CDMFT using the maximum entropy method \cite{Jarrel,Bergeron} (color map). The white solid lines indicate the DFT bands. The Fermi level is at $\omega = 0$. (e) The real part of the CDMFT self-energy on the Matsubara frequency axis. 
	}
	\label{fig2}
\end{figure}

Figure~\ref{fig2} presents the FSs obtained within DMFT and CDMFT. We find from DMFT that the local correlations alone do not affect the FS topology [Fig.~\ref{fig2}(a)]. The size and the shape of three FS pockets obtained from DFT, namely $\alpha$, $\beta$, and $\gamma$ pockets, remain intact. This result is consistent with previous DFT+DMFT studies \cite{shilenko_correlated_2023,lechermann_electronic_2023,qin_high-t_c_2023}.

The IECs, however, significantly modify this picture [Fig.~\ref{fig2}(b)]. While the $\alpha$ pocket remains nearly unchanged, the $\beta$ and the $\gamma$ pockets are largely affected by IECs. The $\beta$ pocket becomes more diamond shaped with spectral weight at the first Brillouin zone (FBZ) boundary being shifted toward the $X$ point. 
We also find redistribution of electron occupations in favor of half-filled $\bar{z}$ and $\underline{z}$ orbitals with $\langle n_{\bar{z}} \rangle = \langle n_{\underline{z}} \rangle  \simeq 0.93$ compared to the DMFT value of $0.85$.
Most interestingly, the $\gamma$ pocket disappears which results in a Lifshitz transition of the FS. Looking at the orbital character of the FS [Fig.~\ref{fig2}(c)] reveals that $x_-$ and $z_-$ (for the $\beta$ pocket around $X$ point) and $z_+$ (for the $\gamma$ pocket around the $M$ point) BA orbitals underlie the FS modification.

More information can be obtained from the momentum-dependent CDMFT spectral function [Fig.~\ref{fig2}(d)]. 
Near the $X$ point, the second lowest band moves upward such that $x_-$ and $z_-$ states get closer to the Fermi level. The flat $z_+$ character at the $M$ point, on the other hand, sinks below the Fermi level, leading to the disappearance of the $\gamma$ pocket.

To further pinpoint the microscopic role of IECs, we investigate the CDMFT self-energy $\Sigma_{l}(i\omega_n)$ where $\omega_n=(2n+1)\pi/T$ is the fermionic Matsubara frequency with $n$ being integer and $l \in \{x_+,z_+,x_-,z_- \}$. 
Without IECs, $\Sigma_{x_+/z_+}(i\omega_n) =  \Sigma_{x_-/z_-}(i\omega_n)$, so IECs are manifested by a difference of the self-energies between the BA orbitals. 
We first find that $\Sigma_{x_+}(i\omega_n) \simeq \Sigma_{x_-}(i\omega_n)$ over the entire frequency range due to small $t^x_{\perp}$ resulting in negligible IECs.

\begin{figure*} [!htbp] 
	\includegraphics[width=1.0\textwidth]{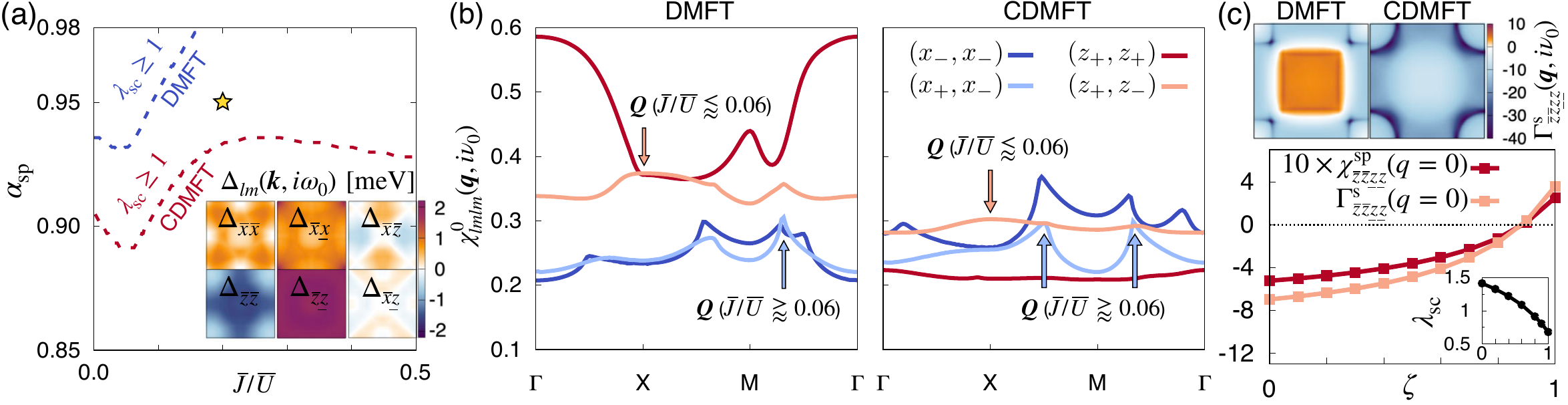}
	\caption{(a) Superconducting phase diagram in the \asp--$\overline{J}/\overline{U}$ space at $T=1/145~\mathrm{eV} \simeq 80$~K. The superconductivity sets in (i.e., $\lambda_\mathrm{sc} \geq 1$) in the regions above the dashed lines; blue for the DMFT and red for the CDMFT. Inset: the CDMFT gap functions in the FBZ for the parameter set marked by the yellow star which corresponds to $\asp=0.95$ and $\overline{J}/\overline{U}=0.2$.  (b) The irreducible susceptibilities at the lowest bosonic frequency $\chi^0_{l m l m}(\bm{q}, i\nu_0)$ calculated using DMFT (left) and CDMFT (right) Green's functions. $\bm{Q}$ and the associated $\bm{\chi}^0(\bm{q}, i\nu_0)$ components are highlighted with colored arrows. (c) Upper panel: the spin-singlet pairing interaction $\Gamma^{\mathrm{s}}_{\bar{z} \bar{z} \underline{z} \underline{z}}(\bm{q},i\nu_0)$ between top and bottom layer $z$ orbitals in the FBZ. Lower panel: $\chi^{\mathrm{sp}}_{\bar{z} \bar{z} \underline{z} \underline{z}}(q=0)$, $\Gamma^{\mathrm{s}}_{\bar{z} \bar{z} \underline{z} \underline{z}}(q=0)$, and $\lambda_\mathrm{sc}$ (inset) as a function of scaling factor $\zeta$ for DMFT $\chi^0_{z_+ z_+  z_+ z_+}(q)$. $\asp=0.95$ and $\overline{J}/\overline{U}=0.2$ for both panels. 
	}
	\label{fig3}
\end{figure*}

In contrast to the $x_{\pm}$ components, large $t^z_{\perp}$ gives rise to strong IECs in the $z_{\pm}$ components. We investigate the real part $\Re \Sigma_l(i\omega_n)$ which modifies the on-site energy level of the orbital $l$; see Fig.~\ref{fig2}(e). See Supplemental Material for the imaginary part \cite{supple}.
We note first that the Hartree-Fock self-energy, $\Re \Sigma_{l}(i\omega_\infty)$, does not modify the FS topology because $\Re \Sigma_{x_\pm }(i\omega_\infty)-\Re \Sigma_{z_\pm }(i\omega_\infty)$ is only about $0.1$~eV and $ \Re \Sigma_{x_+/z_+}(i\omega_\infty) = \Re \Sigma_{x_-/z_-}(i\omega_\infty)$. 

In a low-frequency regime ($\omega_n \ll 10$~eV), however, $\Re \Sigma_{z_+}(i\omega_n)$ is smaller and $\Re \Sigma_{z_-}(i\omega_n)$ is larger than the value at infinite frequency. This, in turn, shifts effectively the onsite energy levels of $z_\pm$ upward ($z_-$) and downward ($z_+$) with respect to their DFT counterparts, thereby enhancing the hybridization gap. 
In fact, this low-energy behavior is the origin of the shifts of spectral weight and concomitant FS change seen in Fig.~\ref{fig2}(b). 
Near the $X$ point $z_-$ has substantial weight in the $\beta$ pocket of the noninteracting FS. Thus the large upturn of $ \Re \Sigma_{z_-}(i\omega_n)$ as $\omega_n \rightarrow 0$ makes an upward shift in energy near the $X$ point, leading to the change of the $\beta$ pocket in CDMFT; see also Supplemental Material \cite{supple}. The physics here bears a close resemblance to that of VO$_2$ in which intersite correlations within dimers promote intradimer singlets with an enhanced hybridization gap  \cite{biermann_dynamical_2005,tomczak_effective_2008,brito_metal-insulator_2016}.

Having analyzed the effects of IECs on the electronic structure, we below investigate how they affect superconductivity. In light of the reported signatures of a spin density wave (SDW) in La$_3$Ni$_2$O$_7$ at ambient pressure \cite{zhang_synthesis_1994,taniguchi_transport_1995,wu_magnetic_2001,ling_neutron_2000,liu_evidence_2022,chen2023evidence,dan2024spindensitywave,chen2024electronic}, it may be natural to consider
spin-fluctuation-mediated pairing. 

A phase transition to the superconducting state occurs when the corresponding pairing susceptibility diverges, which requires numerical evaluation of the pairing vertex $\mathbf{\Gamma}^{\mathrm{s/t}}$ for singlet (s) or triplet (t) Cooper pairs~\cite{bickers_self-consistent_2004,rohringer_diagrammatic_2018} (bold symbols will be used to denote vectors and matrices). The spin and charge susceptibilities ($\bm{\chi}^{\mathrm{sp/ch}}$) 
and the related irreducible vertices ($\mathbf{\Gamma}^{\mathrm{sp/ch}}$) contribute to $\mathbf{\Gamma}^{\mathrm{s/t}}$. Calculating frequency- and momentum-dependent $\mathbf{\Gamma}^{\mathrm{sp/ch}}$ and $\bm{\chi}^{\mathrm{sp/ch}}$, however, is highly nontrivial for multiorbital systems. We thus follow an idea previously employed to study cuprates, ruthenates, and iron-based superconductors \cite{maier_spin_2007,nourafkan_correlation-enhanced_2016,gingras_superconducting_2019,kaser_interorbital_2022}. 
Namely, $\mathbf{\Gamma}^{\mathrm{sp/ch}}$ are parametrized by effective intraorbital Coulomb interaction $\overline{U}$ and Hund's coupling $\overline{J}$, i.e., $\mathbf{\Gamma}^{\mathrm{sp/ch}}  \rightarrow \overline{\mathbf{\Gamma}}^{\mathrm{sp/ch}}(\overline{U},\overline{J})$ (we assume the interorbital value $\overline{U}' = \overline{U}-2\overline{J}$).
The effective vertices $\overline{\mathbf{\Gamma}}^{\mathrm{sp/ch}}(\overline{U},\overline{J})$ are independent of frequency and momentum, see Supplemental Material \cite{supple}.
This leads to the gap equation
\begin{align}
\begin{split}
	\lambda_{\mathrm{sc}} \Delta_{l m}(k) &= -\frac{T}{2N} \sum_{q, l_1 l_2 m_1 m_2 } \Gamma^{\mathrm{s/t}}_{ l l_1 m_1 m}(q) \\ \times &G_{l_1 l_2 } (k-q) G_{m_1 m_2 }(q-k) \Delta_{l_2 m_2 }(k-q),
\end{split}
   \label{gapeq}
\end{align}
where $\lambda_{\mathrm{sc}}$ is the eigenvalue, $G(k)$ the (C)DMFT Green's function, and $\Delta(k)$ the gap function. $k \equiv (\bm{k},i\omega_n)$ and $q \equiv (\bm{q},i\nu_n)$ with $\bm{k}$ and $\bm{q}$ being the crystal momentum and $\nu_n = 2n\pi/T$ the bosonic Matsubara frequency. $N$ indicates the number of $\bm{k}$-points in the FBZ. $\Gamma^{\mathrm{s/t}}_{ l m_1 l_1 m}(q=k-k')$ describe the particle-particle scattering of electrons in orbitals $(l,m)$ with four-momenta $(k,-k)$ to $(l_1,m_1)$ with $(k',-k')$.
The transition to the superconducting state is indicated by the maximum eigenvalue $\lambda_\mathrm{sc}$ reaching unity. Since $\overline{\mathbf{\Gamma}}^{\mathrm{sp/ch}}(\overline{U},\overline{J})$ are more sparse in the original $e_g$ basis than the BA basis, so are the resulting $\bm{\chi}^{\mathrm{sp/ch}}$ and $\mathbf{\Gamma}^{\mathrm{s/t}}$.  We therefore discuss $\bm{\chi}^{\mathrm{sp/ch}}$ and $\mathbf{\Gamma}^{\mathrm{s/t}}$ in the $e_g$ picture.

We find the predominance of singlet over triplet pairings arising from antiferromagnetic (AFM) fluctuations. Figure~\ref{fig3}(a) presents the resulting superconducting phase diagram for the leading singlet channel. Since we cannot pinpoint the precise magnitude of $\overline{U}$ and $\overline{J}$, we scan a range of values. The vertical axis is given by the Stoner enhancement factor \asp which indicates the maximum eigenvalue of $\overline{\mathbf{\Gamma}}^{\mathrm{sp}}(\overline{U} , \overline{J}) \bm{\chi}^0(\bm{q},i\nu_0)$ and gauges the proximity to a magnetic instability. 
Here, $\bm{\chi}^0$ is the irreducible susceptibility, $\chi^0_{l m l' m'}(q)	= -\frac{T}{N}\sum_{k}G_{l l'}(k+q)G_{m' m}(k)$, which is the lowest-order term of the spin susceptibility, $\bm{\chi}^{\mathrm{sp}}=\bm{\chi}^0[\mathbf{1} - \overline{\mathbf{\Gamma}}^{\mathrm{sp}}(\overline{U} , \overline{J})\bm{\chi}^0]^{-1}$. Thus, \asp is determined entirely from $\overline{U}$ and $\overline{J}$, provided $\bm{\chi}^0(\bm{q},i\nu_0)$ is given~\cite{supple}. 
For both DMFT and CDMFT cases, the leading pairing symmetry in the $e_g$-orbital basis is always the intraorbital $s$-wave and interorbital $d_{x^2-y^2}$-wave pairing; see the inset of Fig.~\ref{fig3}(a). Projecting to the noninteracting FS, this pairing corresponds to the $s \pm$-wave where the gap changes sign between the neighboring FS pockets \cite{supple}.
This leading pairing symmetry is in line with many previous studies \cite{zhang_structural_2023,sakakibara_possible_2023,yang_possible_2023,gu_effective_2023,liao_electron_2023,lu_interlayer_2023,liu_spm-wave_2023,oh_type_2023}. The qualitative features of the gap functions remain unchanged over the entire parameter range while $\bm{Q}$ (the crystal momentum at which the maximum eigenvalue of $\overline{\mathbf{\Gamma}}^{\mathrm{sp}} (\overline{U} , \overline{J}) \bm{\chi}^0(\bm{q},i\nu_0)$ emerges) changes from  $\bm{Q} = (\pi,0)$ for $\overline{J}/\overline{U} \lessapprox 0.06$ to an incommensurate wave vector around the $M$ point for $\overline{J}/\overline{U} \gtrapprox 0.06$ for both DMFT and CDMFT as highlighted in Fig.~\ref{fig3}(b); see Supplemental Material for the details \cite{supple}.

The most notable feature of the phase diagram presented in Fig.~\ref{fig3}(a) is the enhanced superconducting instabilities in CDMFT compared to DMFT. 
This result is quite surprising because the $\gamma$ pocket which disappears by IECs within our CDMFT calculation has been argued to drive the spin-fluctuation-mediated superconductivity throughout the literature \cite{zhang_structural_2023,zhang_trends_2023,sakakibara_possible_2023,lechermann_electronic_2023,liu_spm-wave_2023}. It thus raises the question: What is the role of the $\gamma$ pocket in the pairing?

We first find that the $\gamma$ pocket has a ``Janus-faced" role: It hosts both obstructive and supportive magnetic fluctuations for the singlet pairing. This can be, in fact, traced back to the behavior of $\chi^0_{l m l m}(\bm{q},i\nu_0)$ by investigating the $(l, m) = (z_+, z_+)$ and $(z_+, z_-)$ components of the DMFT calculation in Fig.~\ref{fig3}(b). While the $\gamma$ pocket allows for small-$\bm{q}$ particle-hole excitations resulting in the $\bm{q}=(0,0)$ interlayer FM $\chi^0_{z_+ z_+  z_+ z_+}(\bm{q},i\nu_0)$, the $\bm{q} = (\pi, 0)$ nesting between the $\gamma$ and $\beta$ pocket gives rise to AFM $\chi^0_{z_+ z_-  z_+ z_-}(\bm{q},i\nu_0)$. Thus, the $\gamma$ pocket promotes two different competing (i.e., FM vs.~AFM) magnetic channels. Importantly, however, the FM $\chi^0_{z_+ z_+  z_+ z_+}(\bm{q},i\nu_0)$ predominates in DMFT as clearly shown in the left panel of Fig.~\ref{fig3}(b).

The disappearance of the $\gamma$ pocket from the FS due to IECs within CDMFT results in the suppression of both channels, especially the $\chi^0_{z_+ z_+  z_+ z_+}$ component involving solely the $\gamma$ pocket [Fig.~\ref{fig3}(b)]. This change is more apparent from the sign of the pairing interaction.
In the singlet channel, the FM fluctuation is directly manifested by a repulsive (rather than attractive) interaction $\Gamma^{\mathrm{s}}_{\bar{z} \bar{z} \underline{z} \underline{z}}(q=0)$ [upper panel of Fig.~\ref{fig3}(c)], which hinders the singlet Cooper pairing between $\bar{z}$ and $\underline{z}$ orbitals. Quenching of the FM $\chi^0_{z_+ z_+  z_+ z_+}$ as in CDMFT yields the attractive pairing interaction $\Gamma^{\mathrm{s}}_{\bar{z} \bar{z} \underline{z} \underline{z}}$ over the entire FBZ; see Fig.~\ref{fig3}(c).
Hence, the enhanced pairing tendency in CDMFT is mainly attributed to the suppression of this FM channel upon undergoing the Lifshitz transition. 

To further corroborate this argument, we analyze how the DMFT superconducting instabilities are affected by the FM fluctuation by introducing a scaling factor $\zeta$ for $\chi^0_{z_+ z_+  z_+ z_+}(q)$. Namely, $\chi^0_{z_+ z_+  z_+ z_+}(q)$ is rescaled to $\zeta \chi^0_{z_+ z_+  z_+ z_+}(q)$ before constructing $\bm{\chi}^{\mathrm{sp/ch}}$ and $\bm{\Gamma}^\mathrm{s}$. Indeed, as shown in the lower panel of Fig.~\ref{fig3}(c), the interlayer FM spin susceptibility $\chi^\mathrm{sp}_{\bar{z} \bar{z} \underline{z} \underline{z}}(q=0)$ turns AFM with decreasing $\zeta$ followed by an attractive pairing interaction $\Gamma^\mathrm{s}_{\bar{z} \bar{z} \underline{z} \underline{z}}(q=0)$ and an increase of $\lambda_\mathrm{sc}$. See Supplemental Material for additional data \cite{supple}. Note also that since the $\gamma$ pocket is the only FS pocket dispersive along the $k_z$ direction \cite{labollita_electronic_2023}, the disappearance of the $\gamma$ pocket by IECs makes La$_3$Ni$_2$O$_7$ effectively two dimensional.

We now turn to the question of ``Why does superconductivity emerge only in the high-pressure phase?" A useful insight is obtained from a recent experiment which reports that pressure mainly shrinks the out-of-plane Ni-O bond length while the in-plane one is weakly affected \cite{wang2023structure}. Thus, the main effect of pressure can be addressed with the change of $t^z_\perp$ which is sensitive to the out-of-plane Ni-O bond length. Since $t^z_\perp \simeq -0.63$~eV at 29.5~GPa under which superconductivity emerges~\cite{luo_bilayer_2023,gu_effective_2023,sakakibara_possible_2023}, a smaller magnitude of $t^z_\perp$ should correspond to the lower pressure case. In light of this observation, we investigate two ``low pressure" cases, namely $t^z_\perp = -0.45$ and  $t^z_\perp = -0.55$~eV, using CDMFT.

\begin{figure} [!htbp]	 
	\includegraphics[width=0.98\columnwidth]{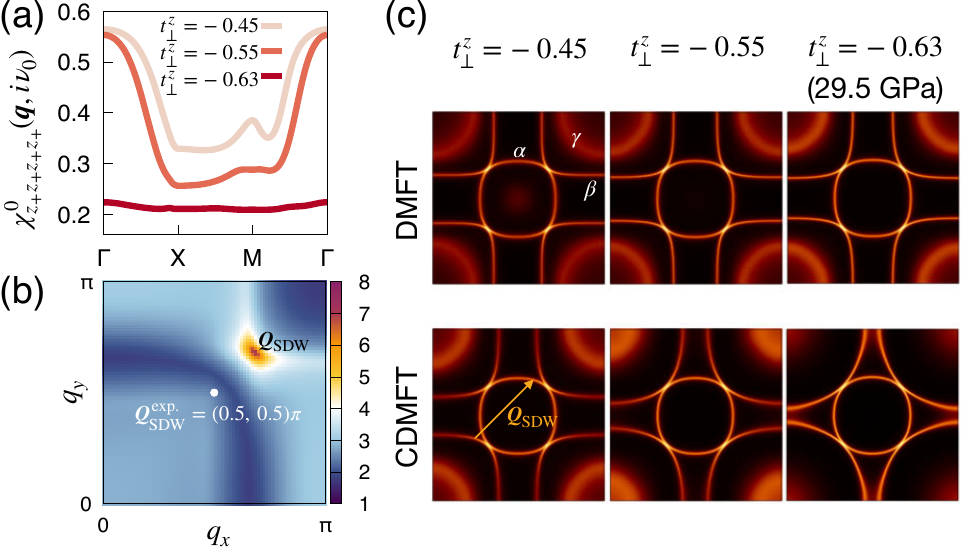}
	\caption{(a) $\chi^0_{z_+ z_+ z_+ z_+}(\bm{q}, i\nu_0)$  calculated using CDMFT Green's functions for three different values of $t^z_\perp$ (in units of eV). (b) $\chi^\mathrm{sp}_\mathrm{max}(\bm{q}, i\nu_0)$ at $\alpha_\mathrm{sp}=0.95$ and $\overline{J}/\overline{U}=0.2$ for $t^z_\perp = -0.45$~$\mathrm{eV}$. (c) FSs obtained from DMFT and  CDMFT for the same values of $t^z_\perp$.  We used $U=3.7$, $J=0.6$~eV, and $U'=U-2J$ for all the cases. $\bm{Q}_\mathrm{SDW}$ in (b) is highlighted with the orange arrow.
	}
	\label{fig4}
\end{figure}

In Fig.~\ref{fig4}(a), we find that large $\chi^0_{z_+ z_+  z_+ z_+}(\bm{q},i\nu_0)$ emerges for the two low-pressure cases, in sharp contrast to the high-pressure result ($t^z_\perp \simeq -0.63$~eV) which we have already noticed in Fig.~\ref{fig3}. Interestingly, this result provides a plausible scenario of why La$_3$Ni$_2$O$_7$  is not superconducting in the low-pressure phase because FM $\chi^0_{z_+ z_+  z_+ z_+}(\bm{q},i\nu_0)$ obstructs the singlet pairing as detailed above. While the $\gamma$ pocket gives rise to strong FM fluctuations, the actual magnetic transition occurs at a finite $\bm{q}$ as shown in Fig.~\ref{fig4}(b) which presents the maximum eigenvalue of the spin susceptibility at each $\bm{q}$, $\chi^\mathrm{sp}_\mathrm{max}(\bm{q}, i\nu_0)$, for $t^z_\perp = -0.45$~eV. $\chi^\mathrm{sp}_\mathrm{max}(\bm{q}, i\nu_0)$ shows a peak at $\bm{q}=\bm{Q}_\mathrm{SDW}$ which is different from but close to the SDW wave vector $\bm{Q}^\mathrm{exp.}_\mathrm{SDW} = (0.5, 0.5)\pi$ reported by experiments \cite{chen2024electronic, dan2024spindensitywave}. See Supplemental Material for further discussion \cite{supple}.

To trace the origin of the difference between the two low-pressure cases and the high-pressure ($t^z_\perp \simeq -0.63$~eV) case, we investigate the CDMFT FSs [lower panels in Fig.~\ref{fig4}(c)].  The strength of IECs is controlled by $|t^z_\perp|$, so the shape of the $\beta$ and the $\gamma$ pockets in CDMFT FS is basically the same as the DMFT FS for the smallest $|t^z_\perp|$ ($t^z_\perp = -0.45$~eV). As $|t^z_\perp|$ increases, however, the $\gamma$ pocket gets suppressed in CDMFT which is consistent with the evolution of $\chi^0_{z_+ z_+ z_+ z_+}(\bm{q}, i\nu_0)$ presented in Fig.~\ref{fig4}(a); see Supplemental Material for further discussion on the microscopic origin, especially on the change of the $\beta$ pocket \cite{supple}. At $t^z_\perp \simeq -0.63$~eV, the FS that we have already seen in Fig.~\ref{fig2}(b) is realized.
In contrast, the FS is almost unaffected by $t^z_\perp$ in DMFT. Hence, it can be seen that the concerted effect of pressure (as modeled via $t^z_\perp$) and IECs induces the Lifshitz transition. This transition quenches the FM channel resulting in an enhancement of the singlet pairing mediated by AFM fluctuations.

We finally discuss implications of the above pressure-induced FS change for the available experimental data. Since the SDW is known to emerge in the ambient-pressure phase \cite{zhang_synthesis_1994,taniguchi_transport_1995,wu_magnetic_2001,ling_neutron_2000,liu_evidence_2022,chen2023evidence,dan2024spindensitywave,chen2024electronic}, direct comparison of our FS for small $|t^z_\perp|$ with experimental FS obtained from angle-resolved photoemission spectroscopy under ambient pressure \cite{yang_orbital-dependent_2023} may be misleading. Also, considering that there is a discrepancy as to whether or not the $\gamma$ pocket crosses the Fermi level in La$_4$Ni$_3$O$_{10}$ even between experiments \cite{dessau2017,du2024correlated}, the same issue may also pertain to La$_3$Ni$_2$O$_7$. Further study is required.
Rather, a tantalizing signature of the Lifshitz transition of FS is seen in the pressure dependence of Hall coefficient ($R_\mathrm{H}$) \cite{zhang_high-temperature_2023,zhou2023evidence}. While the sign of $R_\mathrm{H}$ is positive for the entire pressure range, a drop of $R_\mathrm{H}$ occurs near $\sim 10$~GPa followed by the emergence of superconductivity \cite{zhang_high-temperature_2023,zhou2023evidence}. Since the $\gamma$ pocket is a holelike FS [Fig.~\ref{fig2}(d)] and is destructive for pairing, 
the drop of $R_\mathrm{H}$ and the emergence of superconductivity is quite naturally explained from our Lifshitz transition scenario.

To conclude, we have demonstrated that IECs play a critical role in La$_3$Ni$_2$O$_7$ by inducing a Lifshitz transition. The superconducting instability is found to be enhanced by this transition due to the quenching of the FM fluctuation, which may also explain why superconductivity emerges only under high pressure.


{\it Acknowledgments}. We are grateful to F.~Lechermann and I. Eremin for useful discussion. S.~R. thanks Se Young Park for helpful comments on the charge self-consistency.
This work is supported by the Cluster of Excellence `CUI: Advanced Imaging of Matter' of the Deutsche Forschungsgemeinschaft (DFG) - EXC 2056 - project ID 390715994, by DFG priority program SPP 2244 (WE 5342/5-1 project No. 422707584) and the DFG research unit FOR 5242 (WE 5342/7-1, project No. 449119662).
Calculations were done on the supercomputer Lise at NHR@ZIB as part of the NHR infrastructure under the project hhp00056.



%

\end{document}


\renewcommand{\thepage}{S\arabic{page}}  
\renewcommand{\thesection}{SM\arabic{section}}   
\renewcommand{\thetable}{S\arabic{table}}   
\renewcommand{\thefigure}{S\arabic{figure}}
\renewcommand{\theequation}{S\arabic{equation}}

\renewcommand{\citenumfont}[1]{S#1}
\renewcommand{\bibnumfmt}[1]{[S#1]}

\def\tcb{\textcolor{blue}}
\def\tcr{\textcolor{red}}
\def\tcg{\textcolor{green}}
\def\tcc{\textcolor{cyan}}

\onecolumngrid

\title{Supplemental Material for \\ ``Quenched pair breaking by interlayer correlations as a key to superconductivity in La$_3$Ni$_2$O$_7$"}

\author{Siheon Ryee}
\email{sryee@physnet.uni-hamburg.de} 
\affiliation{I. Institute of Theoretical Physics, University of Hamburg, Notkestrasse 9, 22607 Hamburg, Germany}

\author{Niklas Witt}
\affiliation{I. Institute of Theoretical Physics, University of Hamburg, Notkestrasse 9, 22607 Hamburg, Germany}
\affiliation{The Hamburg Centre for Ultrafast Imaging, Luruper Chaussee 149, 22761 Hamburg, Germany}

\author{Tim O. Wehling}
\affiliation{I. Institute of Theoretical Physics, University of Hamburg, Notkestrasse 9, 22607 Hamburg, Germany}
\affiliation{The Hamburg Centre for Ultrafast Imaging, Luruper Chaussee 149, 22761 Hamburg, Germany}


\maketitle
\tableofcontents
\hypersetup{linkcolor=red}

\newpage

\section{Microscopic model} \label{SM1}
We consider a Hamiltonian on the bilayer square lattice: $\mathcal{H}=H_0 + H_\mathrm{int}$. Here, $H_0$ is a tight-binding term describing the noninteracting band structure, which reads
\begin{align} \label{eqS1}
H_0 &= \sum_{il, jm, \sigma} t_{il,jm} d^{\dagger}_{il \sigma} d_{jm, \sigma},
\end{align}
where $d^{\dagger}_{il, \sigma}$ ($d_{j m, \sigma}$) is the electron creation (annihilation) operator for site $i,j$ of the bilayer square lattice, spin $\sigma \in \{ \uparrow, \downarrow \}$, and orbital $l, m \in \{ \bar{\eta}, \underline{\eta} \}$ where $\bar{\eta}$ and $\underline{\eta}$ represent Ni-$e_g$ orbitals $\eta \in  \{ x^2-y^2, z^2 \}$ in the top and bottom layers, respectively. We simply denote $x^2-y^2$ by $x$ and $z^2$ by $z$.
$\{ t_{il,jm} \}$ are the hopping amplitudes for which we use values obtained from the maximally localized Wannier function (MLWF) description of the DFT result for La$_3$Ni$_2$O$_7$ under high pressure (29.5~GPa) \cite{gu_effective_2023}. Note in particular that $t^\eta_{\perp} \equiv t_{i \bar{\eta},i \underline{\eta}} = t_{i \underline{\eta},i \bar{\eta}} $ with $t^x_{\perp}=-0.049$~eV and $t^z_{\perp}=-0.628$~eV \cite{gu_effective_2023}. The band dispersion of $H_0$ is presented in Fig.~2(c) in the main text.
$H_\mathrm{int}$ is the onsite interaction term given by the standard Kanamori form
\begin{align}
	H_\mathrm{int} = \frac{1}{2}\sum_{i, l m' m l', \sigma \sigma'} \mathcal{U}^{\sigma \sigma'}_{l m' m l'} d^{\dagger}_{i l \sigma} d^{\dagger}_{i l' \sigma'} d_{i m' \sigma'} d_{i m \sigma},
	\label{Hint}
\end{align}
where $l,l',m,m' \in \{ \bar{x}, \bar{z},  \underline{x}, \underline{z}  \}$. $\mathcal{U}^{\sigma \sigma'}_{l m' m l'}$ is nonzero only for $U \equiv \mathcal{U}^{\sigma \sigma}_{\bar{\eta} \bar{\eta} \bar{\eta} \bar{\eta}} = \mathcal{U}^{\sigma -\sigma}_{\bar{\eta} \bar{\eta} \bar{\eta} \bar{\eta}} =\mathcal{U}^{\sigma \sigma}_{\underline{\eta} \underline{\eta} \underline{\eta} \underline{\eta}} =\mathcal{U}^{\sigma -\sigma}_{\underline{\eta} \underline{\eta} \underline{\eta} \underline{\eta}} $, $U' \equiv \mathcal{U}^{\sigma \sigma}_{\bar{\eta} \bar{\eta}' \bar{\eta} \bar{\eta}' } = \mathcal{U}^{\sigma -\sigma}_{\bar{\eta} \bar{\eta}' \bar{\eta} \bar{\eta}' } = \mathcal{U}^{\sigma \sigma}_{\underline{\eta} \underline{\eta}' \underline{\eta}  \underline{\eta}' } = \mathcal{U}^{\sigma -\sigma}_{\underline{\eta} \underline{\eta}' \underline{\eta}  \underline{\eta}' } $, and $J  \equiv  \mathcal{U}^{\sigma \sigma}_{\bar{\eta} \bar{\eta}' \bar{\eta}' \bar{\eta}} =  \mathcal{U}^{\sigma -\sigma}_{\bar{\eta} \bar{\eta}' \bar{\eta}' \bar{\eta}} = \mathcal{U}^{\sigma \sigma}_{\bar{\eta} \bar{\eta} \bar{\eta}' \bar{\eta}'}=\mathcal{U}^{\sigma -\sigma}_{\bar{\eta} \bar{\eta} \bar{\eta}' \bar{\eta}'} =  \mathcal{U}^{\sigma \sigma}_{\underline{\eta} \underline{\eta}' \underline{\eta}' \underline{\eta}} = \mathcal{U}^{\sigma -\sigma}_{\underline{\eta} \underline{\eta}' \underline{\eta}' \underline{\eta}}=  \mathcal{U}^{\sigma \sigma}_{\underline{\eta} \underline{\eta} \underline{\eta}' \underline{\eta}'} = \mathcal{U}^{\sigma -\sigma}_{\underline{\eta} \underline{\eta} \underline{\eta}' \underline{\eta}'}$ ($\eta \neq \eta'$).  We take $U=3.7$, $J =0.6$~eV, and $U'=U-2J$ by taking {\it ab initio} estimates for the $e_g$ MLWF model \cite{christiansson_correlated_2023}. 

To mitigate the Monte Carlo sign problem of the two-site cluster impurity, we solve the model in a bonding-antibonding (BA) basis in which the electron annhilation operator $\widetilde{d}_{i\eta_\pm \sigma}$ is defined as the symmetric ($+$) and antisymmetric ($-$) combinations of the top and bottom layer $e_g$ orbital operators:
\begin{align}
\begin{pmatrix}
	\widetilde{d}_{ix_+ \sigma} \\
	\widetilde{d}_{iz_+ \sigma} \\	
	\widetilde{d}_{ix_- \sigma} \\
	\widetilde{d}_{iz_- \sigma} 
\end{pmatrix} 
= \mathbf{A} 
\begin{pmatrix}
	d_{i\bar{x} \sigma} \\
	d_{i\bar{z} \sigma} \\	
	d_{i\underline{x}\sigma} \\
	d_{i\underline{z} \sigma} 
\end{pmatrix}
= \frac{1}{\sqrt{2}} \begin{pmatrix}  1 & 0 & 1 & 0 \\ 0 & 1 & 0 & 1 \\  1 & 0 & -1 & 0 \\ 0 & 1 & 0 & -1 \end{pmatrix} 
\begin{pmatrix}
	d_{i\bar{x} \sigma} \\
	d_{i\bar{z} \sigma} \\	
	d_{i\underline{x}\sigma} \\
	d_{i\underline{z} \sigma} 
\end{pmatrix}, ~\mathrm{where}~\mathbf{A} = \frac{1}{\sqrt{2}} \begin{pmatrix}  1 & 0 & 1 & 0 \\ 0 & 1 & 0 & 1 \\  1 & 0 & -1 & 0 \\ 0 & 1 & 0 & -1 \end{pmatrix}.
\end{align}
Under this basis transformation, $H_0$ and $H_\mathrm{int}$ are rewritten as
\begin{align}
&H_0 = \sum_{il, jm, \sigma} \Big(  \sum_{l_1, m_1}  t_{il_1,jm_1}A_{l_1 l}A^*_{m_1 m} \Big) \widetilde{d}^{\dagger}_{il \sigma} \widetilde{d}_{jm, \sigma}, \\
&H_\mathrm{int} = \frac{1}{2}\sum_{i,l m' m l', \sigma \sigma'}   \Big( \sum_{l_1 m_2 m_1 l_2}  \mathcal{U}^{\sigma \sigma'}_{l_1 m_2 m_1 l_2} A_{l_1 l}A_{l_2 l'}A^*_{m_1 m} A^*_{m_2 m'} \Big) \widetilde{d}^{\dagger}_{i l \sigma} \widetilde{d}^{\dagger}_{i l' \sigma'} \widetilde{d}_{i m' \sigma'} \widetilde{d}_{i m \sigma}.
\end{align}
The transformed Coulomb interaction tensor to the BA basis, namely $\widetilde{\mathcal{U}}^{\sigma \sigma'}_{l m' m l'}=\sum_{l_1 m_2 m_1 l_2}  \mathcal{U}^{\sigma \sigma'}_{l_1 m_2 m_1 l_2} A_{l_1 l}A_{l_2 l'}A^*_{m_1 m} A^*_{m_2 m'}$, has in general nonzero four-index elements. While the four-index terms cause the Monte Carlo sign problem, it is actually largely alleviated in the BA basis compared to the $e_g$ basis thanks to the orbital-diagonal hybridization function.

\section{Imaginary part of the CDMFT self-energy}

\begin{figure} [!htbp] 
	\includegraphics[width=0.3\textwidth]{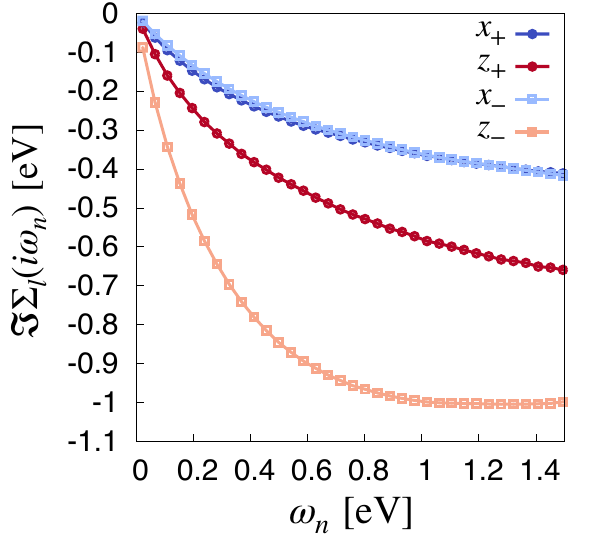}
	\caption{The imaginary part of the CDMFT self-energy on the Matsubara frequency axis. }
	\label{sfig_imag_sigma}
\end{figure}

The imaginary part of the CDMFT self-energy, $\Im \Sigma_{l}(i\omega_n)$ is presented in Fig.~\ref{sfig_imag_sigma}. Noticeable is the strong orbital dependence of $\Im \Sigma_{l}(i\omega_n)$: the $z_{\pm}$ components are more correlated than the $x_{\pm}$, i.e., $|\Im \Sigma_{z_{\pm}}(i\omega_n)| > |\Im \Sigma_{x_{\pm}}(i\omega_n)|$. The result is qualitatively consistent with the available experimental data on the ambient-pressure structure of La$_3$Ni$_2$O$_7$ \cite{yang_orbital-dependent_2023,liu_electronic_2023}. The origin is traced back to the resulting electron occupation of the $\bar{z}$ and $\underline{z}$ orbitals ($\langle n_{\bar{z}}  \rangle= \langle n_{\underline{z}} \rangle  \simeq 0.93$) being much closer to half filling than the $\bar{x}$ and $\underline{x}$ ($\langle n_{\bar{x}} \rangle = \langle n_{\underline{x}} \rangle  \simeq 0.57$). The IECs between $\bar{z}$ and $\underline{z}$ orbitals further differentiate $z_+$ and $z_-$, namely $|\Im \Sigma_{z_-}(i\omega_n)| > |\Im \Sigma_{z_+}(i\omega_n)|$, leading to a larger mass enhancement $m^*/m_\mathrm{DFT}$ ($m_\mathrm{DFT}$ is the bare DFT mass) in the $z_-$ component. The mass enhancements directly extracted from $\Im \Sigma_l(i\omega_n)$ using fourth-order polynomial fitting \cite{mravlje_coherence-incoherence_2011,ryee_hund_2021,ryee_frozen_2023} are $m^*/m_\mathrm{DFT} \simeq 2.8$ for $z_+$ and $5.1$ for $z_-$.  On the other hand, $m^*/m_\mathrm{DFT} \approx 2$ for both weaker-correlated $x_+$ and $x_-$ orbitals.

\section{Influence of different onsite interaction parameters on the Fermi surface}
We have used {\it ab initio} estimate of $U$ ($U=3.7$~eV) and $J$ ($J=0.6$~eV) obtained from the constrained random phase approximation (cRPA) \cite{christiansson_correlated_2023} for our presentation in the main text. Since we have neglected i) frequency dependence of $U$ and $J$ by using their static limits and ii) spatially long-ranged Coulomb tails by taking only onsite values, it would be necessary to check how the low-energy electronic structure is affected by different values of $U$ and $J$. Here, we change $U$ and keep the ratio $J/U$ fixed.

\begin{figure} [!htbp] 
	\includegraphics[width=0.8\textwidth]{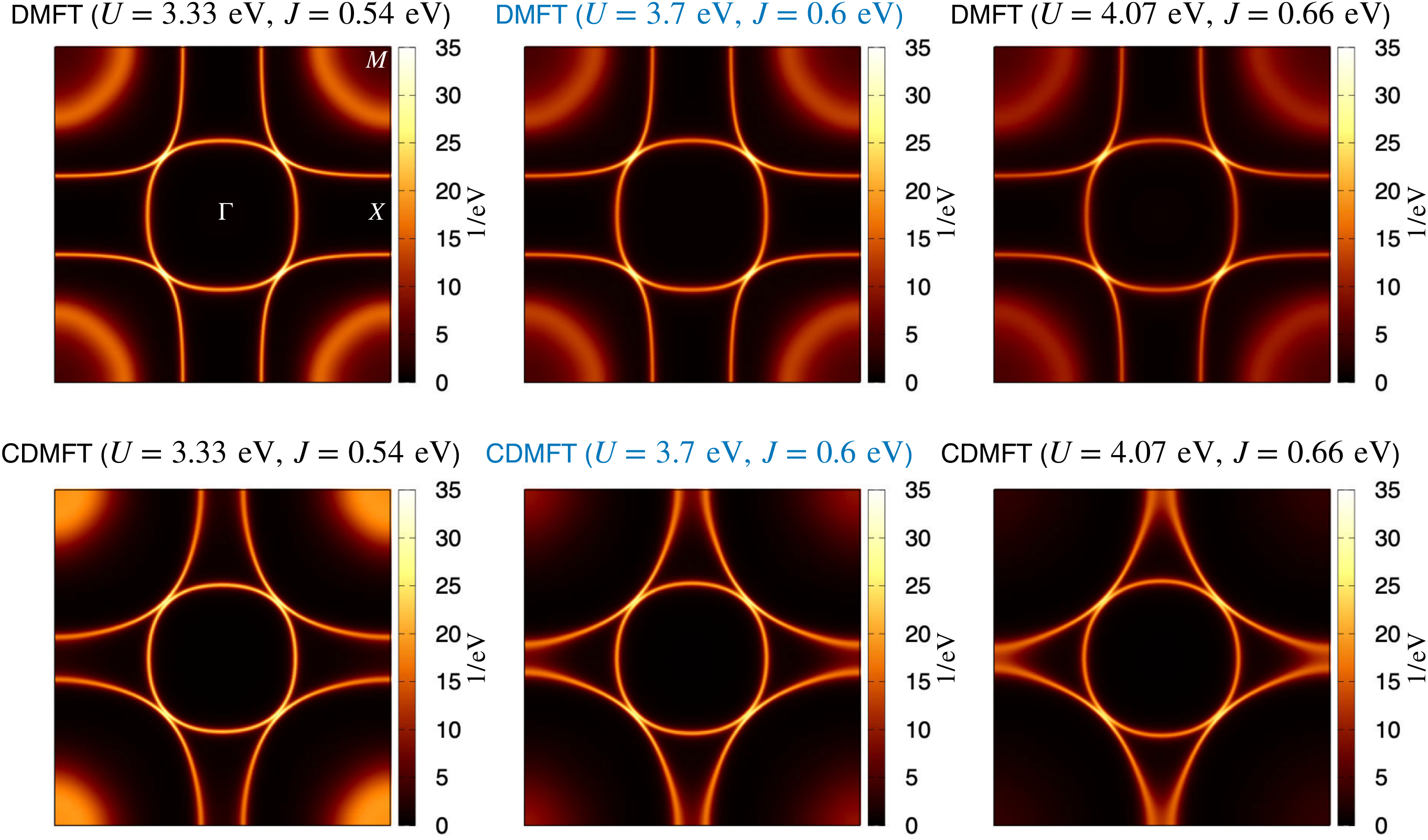}
	\caption{The calculated FSs by DMFT (top panels) and CDMFT (bottom panel) for different interaction parameters. The FSs are approximated to $-\sum_{l m}\delta_{l m}\Im G_{l m}(\bm{k},i\omega_0)$ where $l, m \in \{x_+,z_+,x_-,z_- \}$. The FSs in the middle panels are obtained from the {\it ab initio} interaction parameters as discussed in the main text.  $U'=U-2J$ for all the sets.}
	\label{sfig_FS}
\end{figure}

We present the calculated FSs in Fig.~\ref{sfig_FS}. As is expected, the strength of IECs increases with $U$ and $J$. Namely, the $\gamma$ pocket at the $M$ point gets suppressed and the $\beta$ pocket becomes more diamond-shaped with increasing $U$ and $J$ in CDMFT results. On the other hand, we find no appreciable change within DMFT in which only onsite electronic correlations are taken into account. 
In any case, we find that in a reasonable range around the cRPA estimates IECs significantly modify the FS making the $\gamma$ pocket largely suppressed.

\section{Influence of interlayer density-density interaction $V$}

Using the BA basis for CDMFT calculations, the nearest-neighbor interlayer interactions can be addressed on the same footing as onsite interactions ($U$ and $J$) without invoking further computational complexity. In light of this, we here consider the effects of density-density interaction $V$ within CDMFT by introducing an additional term $H_V$ to our Hamiltonian. $H_V$ reads
\begin{align}
\begin{split}
	H_V &= \sum^{i\sigma \sigma'}_{\eta, \eta' \in \{x,z\} } V d^\dagger_{i \bar{\eta} \sigma} d^\dagger_{i \underline{\eta'} \sigma'} d_{i \underline{\eta'}\sigma'} d_{i \bar{\eta} \sigma} \quad( e_g~\mathrm{basis})\\ 
	&= \sum^{i\sigma \sigma'}_{l, m \in \{x_+,z_+,x_-,z_-\} }  \Big( \sum_{\eta, \eta' \in \{x,z\} }  V  A_{\bar{\eta} l} A_{\underline{\eta'} m}A^*_{\bar{\eta} l} A^*_{\underline{\eta'} m} \Big) \widetilde{d}^{\dagger}_{i l \sigma} \widetilde{d}^{\dagger}_{i m \sigma'} \widetilde{d}_{i m \sigma'} \widetilde{d}_{i l \sigma}  \quad ~(\mathrm{BA ~basis}).
\end{split}
\end{align}
We use the cRPA estimate of $V$: $V=0.5$~eV \cite{christiansson_correlated_2023}.

\begin{figure} [!htbp] 
	\includegraphics[width=1.0\textwidth]{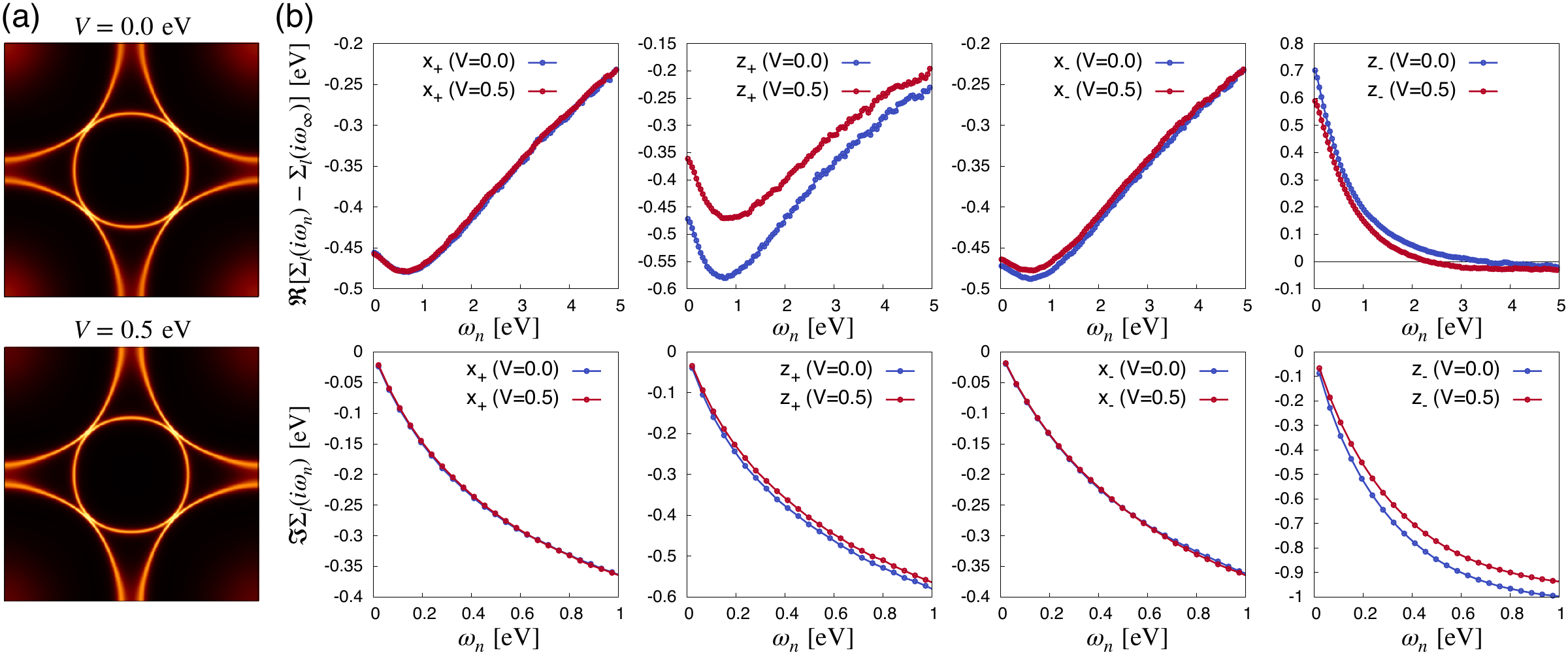}
	\caption{(a) The calculated FSs by CDMFT for $ V=0 $ (top) and $ V=0.5 $~eV (bottom). (b) The real (top) and imaginary (bottom) parts of the CDMFT self-energies. $U=3.7$~eV, $J=0.6$~eV, and $U'=U-2J=2.5$~eV.}
	\label{sfig_V}
\end{figure}

Figure~\ref{sfig_V} presents the FS and the self-energy calculated by CDMFT with $V=0.5$~eV. For comparison, we also show the results with $V=0$. We first find that the FS is basically unaffected by $V$, although $V$ slightly suppresses the $\gamma$ pocket at the $M$ point. This minor difference can be traced back to the small difference in the real part of the self-energy [upper panels in Fig.~\ref{sfig_V}(b)]. For all orbital characters, the effect of $V$ induces at most a $\sim 0.1$~eV decrease in the magnitude of the real part. Note that for comparison of the two cases on the same footing, we subtracted the infinite-frequency Hartree-Fock self-energy from the real part. We also find basically negligible modifications in the imaginary part of the self-energy [lower panels in Fig.~\ref{sfig_V}(b)], albeit them being systematically decreased in magnitude by $V$. We can ascribe these suppressed real and imaginary parts of the self-energy in the finite $V$ case to the generic effect of $V$ which reduces the ``effective" onsite interaction strength by screening \cite{ayral_screening_2013,schuler_optimal_2013,van_loon_capturing_2016,nilsson_multitier_2017,ryee_nonlocal_2020}. 
At any rate, these small changes in the FS and the self-energy legitimately allow us to employ only onsite interactions as we did in the present paper.

\section{Influence of Hund's coupling $J$}
\begin{figure} [!htbp] 
	\includegraphics[width=1.0\textwidth]{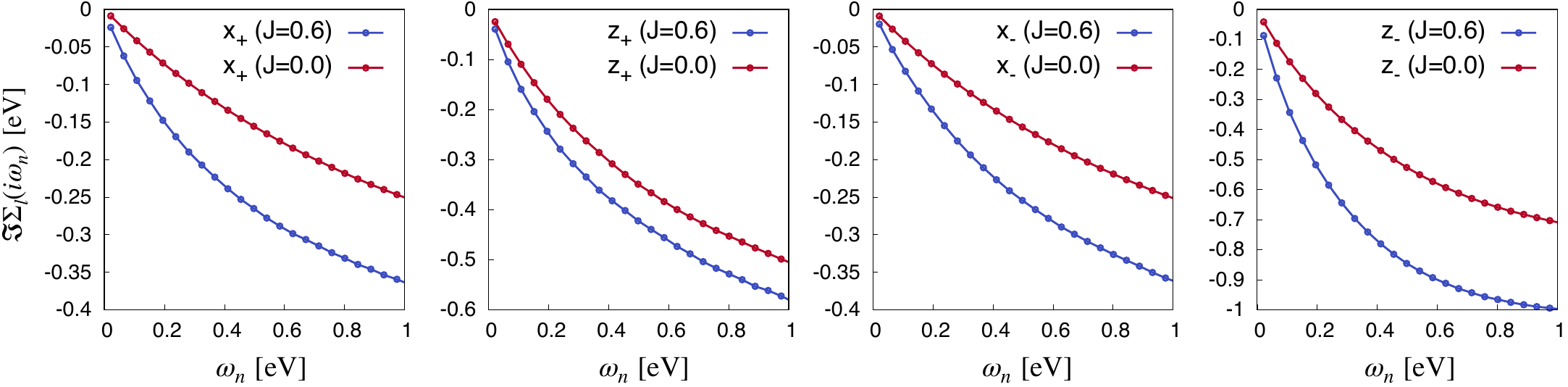}
	\caption{The imaginary parts of the CDMFT self-energies for $J=0.6$~eV (blue) and $J=0.0$ (red). $U=3.7$~eV and $U'=U-2J$ for both cases.}
	\label{sfig_J}
\end{figure}

Since La$_3$Ni$_2$O$_7$ realizes a multiorbital system, it would be informative to investigate how Hund's coupling $J$ affects the low-energy physics. Figure~\ref{sfig_J} presents the effect of $J$ on the imaginary part of the CDMFT self-energy. Interestingly, the magnitude of $\Im \Sigma_l (i\omega_n)$ increases for all the orbital character once we turn on $J$. Thus, the system becomes more correlated by $J$, which in this sense can be identified as a "Hund's metal" \cite{georges_strong_2013}. This is because the formation of large local spin moments promoted by Hund's coupling suppresses Kondo screening, thereby prohibiting the emergence of long-lived quasiparticles \cite{nevidomskyy_kondo_2009,yin_fractional_2012,georges_strong_2013,aron_analytic_2015,ryee_switching_2023,ryee_frozen_2023}.
We also find a strong orbital dependence of $\Im \Sigma_{l}(i\omega_n)$ for both $J=0$ and $J=0.6$~eV (the {\it ab initio} estimate) due to the fact that the electron occupation of the $\bar{z}$ and $\underline{z}$ orbitals is much closer to half filling than that of the $\bar{x}$ and $\underline{x}$ orbitals in both cases. 
The redistribution of electron filling by turning on $J$ is found to be almost negligible in this system. 

\section{Mimicking charge-self-consistent DFT+(C)DMFT}
In this paper, we have addressed the (C)DMFT results to the {\it ab initio} lattice model derived from DFT as defined in Sec.~\ref{SM1}. We here argue that charge self-consistency in the spirit of DFT+(C)DMFT may not affect the findings of our paper.

\begin{figure} [!htbp] 
	\includegraphics[width=1.0\textwidth]{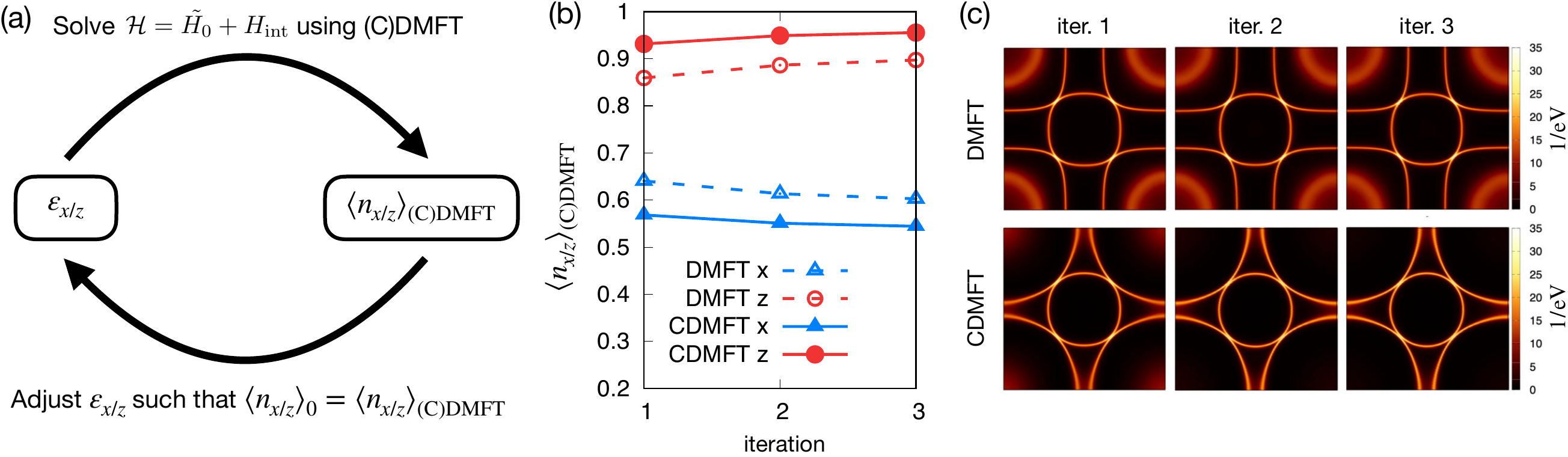}
	\caption{(a) Schematic of mimicking DFT+(C)DMFT charge self-consistency by adjusting the on-site energy levels ($\varepsilon_{x/z}$) for top and bottom layer $x$ and $z$ orbitals. (b) The (C)DMFT electron occupation $\langle n_{x/z} \rangle_\mathrm{(C)DMFT}$ as a function of iteration steps. (c) The resulting (C)DMFT FSs for each iteration step. The FSs in Iteration 1 correspond to the (C)DMFT results presented throughout this paper.}
	\label{sfig_CSC}
\end{figure}

To mimic the charge self-consistency in the low-energy lattice model $\mathcal{H}=H_0 + H_\mathrm{int}$, we here resort to a strategy by noting that a redistribution of electrons by (C)DMFT modifies a charge-density profile $\rho(\bm{r})$ and the resulting local electron occupation, from which the subsequent DFT would result in new on-site energy levels of $x$ and $z$ orbitals for the kinetic part $H_0$. In short, we iterate (C)DMFT calculations for the updated $H_0$, which we call $\tilde{H_0}$ in which the on-site energy levels ($\varepsilon_{x/z}$) are adjusted such that the noninteracting electron occupation ($\langle n_{x/z} \rangle_0$) is equal to the self-consistently determined (C)DMFT occupation ($\langle n_{x/z} \rangle_\mathrm{(C)DMFT}$) of the previous iteration; see Fig.~\ref{sfig_CSC}(a). 

Figure~\ref{sfig_CSC}(b) presents $\langle n_{x/z} \rangle_\mathrm{(C)DMFT}$ as a function of the iteration step. Note that iteration 1 corresponds to the (C)DMFT result which we have discussed throughout the paper. We find that $\langle n_{x/z} \rangle_\mathrm{(C)DMFT}$ almost converges at the third iteration, albeit $\langle n_{x/z} \rangle_\mathrm{(C)DMFT}$ does not change significantly during the iteration. Most importantly, FS remains almost unchanged [Fig.~\ref{sfig_CSC}(c)], which validates (C)DMFT approach employed in this paper. Indeed, our DMFT FS is consistent with the charge-self-consistent DFT+DMFT results reported in Refs.~\cite{shilenko_correlated_2023,lechermann_electronic_2023} .

\section{Why the $\beta$ Fermi-surface pocket becomes more diamond-shaped with pressure}

To model lower pressure cases, we look at a recent experimental study on the evolution of Ni-O bond length as a function of pressure \cite{wang2023structure}. Here, it is reported that pressure mainly shrinks the out-of-plane Ni-O bond length while the in-plane one is very weakly affected \cite{wang2023structure}. Thus, the main effect of pressure can be addressed with the change in the magnitude of $t^z_\perp$ which is sensitive to the out-of-plane Ni-O bond length. Since $t^z_\perp \simeq -0.63$~eV at 29.5~GPa under which superconductivity emerges, a smaller magnitude of $t^z_\perp$ should correspond to the lower pressure condition. In light of this observation, we investigate two different ``low pressure" cases, namely $t^z_\perp = -0.45$~eV and  $t^z_\perp = -0.55$~eV, using CDMFT as well as DMFT.

\begin{figure} [!htbp] 
	\includegraphics[width=1.0\textwidth]{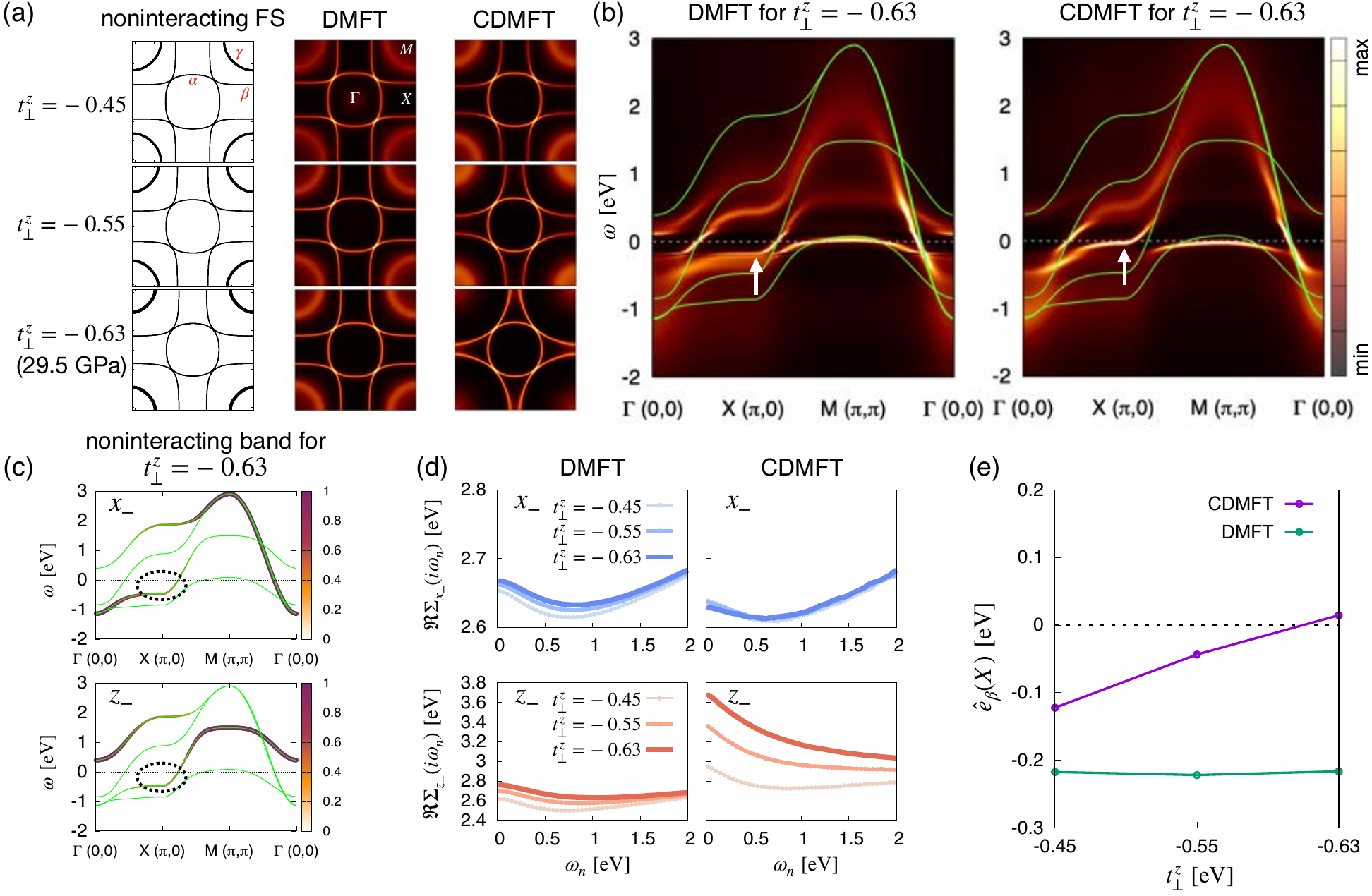}
	\caption{(a) Fermi surfaces obtained from the noninteracting Hamiltonian (left panels), DMFT (middle panels), and CDMFT (right panels) for the three different values of $t^z_\perp$ (in units of eV).  (b) The momentum-dependent spectral function obtained from DMFT (left panel) and CDMFT (right panel) for $t^z_\perp=-0.63$~eV. The green solid lines indicate the DFT bands. The Fermi level is at $\omega = 0$.	(c) The weight of $x_-$ (top panel) and $z_-$ (bottom panel) orbitals in the DFT band structure at the high-pressure phase where $t^z_\perp=-0.63$~eV.  (d) The real part of the DMFT and CDMFT self-energies for $x_-$ (top panels) and $z_-$ (bottom panels) on the Matsubara frequency axis for the same values of $t^z_\perp$. (e) The calculated $\hat{e}_\beta(X)$ according to Eq.~(\ref{eqR8}).
	}
	\label{sfig_pressure}
\end{figure}

Figure~\ref{sfig_pressure}(a) presents Fermi surfaces (FSs) for three different values of $t^z_\perp$. DFT-like noninteracting FS remains almost unchanged albeit the $\gamma$ pocket gets slightly suppressed as $|t^z_\perp|$ increases (or, equivalently as the applied pressure increases). 
In CDMFT, FS is also basically the same as the noninteracting FS for the lowest pressure case in which $t^z_\perp=-0.45$~eV.  Interestingly, however, as $|t^z_\perp|$ increases, the $\beta$ pocket becomes more diamond-shaped in CDMFT.
The shape change of the $\beta$ FS pocket in CDMFT is traced back to the upward energy shift of the spectral weight near the $X$ point of the Brillouin zone in the CDMFT spectral function compared to the DMFT as highlighted with white arrows in Fig.~\ref{sfig_pressure}(b). Furthermore, the shape change of the $\beta$ FS pocket with $|t^z_\perp|$ presented in Fig.~\ref{sfig_pressure}(a) implies that the spectral weight near the $X$ point gets pushed upward in CDMFT with $|t^z_\perp|$.

The reason for such change of the $\beta$ pocket is two-fold: i) the $\beta$ pocket contains the $z_-$ orbital character near the $X$ point as highlighted with black dotted ovals in Fig.~\ref{sfig_pressure}(c) and ii) the real part of the CDMFT $z_-$-orbital self-energy $\Re \Sigma_{z_-}$ increases with pressure as presented in Fig.~\ref{sfig_pressure}(d).
For a more quantitative understanding of this argument, let us now investigate in detail how the energy level at the $X$ point is modified in DMFT and CDMFT. To this end, we first write down the interacting Green's function which reads
\begin{align} \label{eqR1}
	\bm{G^{-1}}(\bm{k},\omega) = (\omega + \mu){\bf I} - \bm{H_0}(\bm{k}) - \bm{\Sigma}(\omega),
\end{align}
where $\omega$ is real frequency, $\bm{k}$ the crystal momentum, $\mu$ a chemical potential. A bold capital letter represents a $4 \times 4$ matrix in the space of BA orbital $l$ ($l \in \{x_+,z_+,x_-,z_- \}$).  $\bf{I}$ is the identity matrix. $\bm{H_0}(\bm{k})$ is a one-particle Hamiltonian derived from DFT, and $\bm{\Sigma}(\omega)$ a self-energy calculated by DMFT or CDMFT. We are interested in how the poles of $\bm{G}(\bm{k},\omega)$ are shifted by electronic correlations, so it is useful to investigate the quasiparticle energies which are the solutions $\omega(\bm{k})$ of
\begin{align} \label{eqR2}
	\mathrm{Det}[ (\omega + \mu){ \bf I} - \bm{H_0}(\bm{k}) - \Re\bm{\Sigma}(\omega) ] = 0.
\end{align}
Near the Fermi level ($\omega =0$), $\Re\bm{\Sigma}(\omega)$ can be expanded to the linear order in $\omega$: 
\begin{align} \label{eqR3}
	\Re\bm{\Sigma}(\omega) \simeq \Re\bm{\Sigma}(0) + \omega({\bf I} - \bm{ Z}^{-1}) \simeq \Re\bm{\Sigma}(i\omega_0) + \omega({\bf I} - \bm{ Z}^{-1}),
\end{align}
where $\bm{Z}^{-1} = ({\bf I} - \partial \Re \bm{\Sigma} (\omega) / \partial \omega )|_{\omega=0} = ({\bf I} - \partial \Im \bm{\Sigma} (i\omega_n) / \partial \omega_n )|_{\omega_n=0}$ is the inverse of the quasiparticle weight $\bm{Z}$. In practice, we evaluate $\bm{Z}$ using $\Im \bm{\Sigma} (i\omega_n)$ defined in the Matsubara frequency space $\omega_n$ by employing the fourth-order polynomial fitting  \cite{mravlje_coherence-incoherence_2011,ryee_hund_2021,ryee_frozen_2023}. $\Re\bm{\Sigma}(i\omega_0)$ is the real-part of the self-energy at the lowest Matsubara frequency, which approximates $\Re\bm{\Sigma}(\omega=0)$. 
We then express $\bm{Z}^{-1}$ as a symmetrical product:
\begin{align} \label{eqR4}
    \bm{Z}^{-1} = \bm{Z}^{-1/2} \bm{Z}^{-1/2}.
\end{align}
By plugging Eq.~(\ref{eqR3}) in Eq.~(\ref{eqR2}) we arrive at
\begin{align} \label{eqR5}
\mathrm{Det}[ \omega {\bf I} - \bm{\sqrt{Z}} \{ \bm{H_0}(\bm{k}) -\mu {\bf I} + \Re\bm{\Sigma}(i\omega_0)\} \bm{\sqrt{Z}} ] = 0.
\end{align}
One can identify a Hermitian matrix $\bm{\sqrt{Z}} \{ \bm{H_0}(\bm{k}) -\mu {\bf I} + \Re\bm{\Sigma}(i\omega_0)\} \bm{\sqrt{Z}} $ as the quasiparticle Hamiltonian whose eigenvalues at each $\bm{k}$ correspond to the quasiparticle energies. A useful insight can be obtained from the change of basis of Eq.~(\ref{eqR5}) to the ``band basis" by which $\bm{H_0}(\bm{k})$ becomes diagonal:
\begin{align}  \label{eqR6}
	\bm{\sqrt{Z}} \{ \bm{H_0}(\bm{k}) -\mu {\bf I} + \Re\bm{\Sigma}(i\omega_0)\} \bm{\sqrt{Z}} \xrightarrow{
		\mathrm{band~basis}} \hat{\bm{\sqrt{Z}}} \{\hat{\bm{H_0}}(\bm{k}) - \mu \hat{\bf I} +  \Re\hat{\bm{\Sigma}}(i\omega_0)  \} \hat{\bm{\sqrt{Z}}} ,
\end{align}
where bold capital letters with hat represent $4 \times 4$ matrices in the band basis. $\hat{\bm{H_0}}(\bm{k}) $ is the diagonal matrix whose elements $e_{n}(\bm{k})$ ($n$: band index) are the band energies. 
We find that off-diagonal elements of $\hat{\bm{\sqrt{Z}}}$ and $\Re\hat{\bm{\Sigma}}(i\omega_0)$ are much smaller than the diagonal elements. Thus the band energy $e_{n}(\bm{k})$ approximately turns into a quasiparticle energy $\hat{e}_{n}(\bm{k})$:
\begin{align} \label{eqR7}
	e_{n}(\bm{k}) \rightarrow \hat{e}_{n}(\bm{k}) &\equiv  \hat{\bm{\sqrt{Z}}}_{nn} \{ e_{n}(\bm{k}) - \mu + \Re\hat{\bm{\Sigma}}(i\omega_0)_{nn} \}  \hat{\bm{\sqrt{Z}}}_{nn}, 
\end{align}
where $\hat{\bm{\sqrt{Z}}}_{nn}$ and $\Re\hat{\bm{\Sigma}}(i\omega_0)_{nn}$ are the $n$-th diagonal elements, respectively, of $\hat{\bm{\sqrt{Z}}}$ and $\Re\hat{\bm{\Sigma}}(i\omega_0)$.
Equation~(\ref{eqR7}) allows us to trace how $e_{n}(\bm{k})$ is shifted in energy due to electronic correlations. Evaluating Eq.~(\ref{eqR7}) at $\bm{k}=X$ for the $\beta$ pocket ($n=\beta$) yields the approximate quasiparticle energy $\hat{e}_{\beta}(X)$ which reads
\begin{align} \label{eqR8}
\begin{split}
    \hat{e}_{\beta}(X) = & \{ |c_{x_-}|^2 \sqrt{Z}_{x_-} + |c_{z_-}|^2 \sqrt{Z}_{z-} \} \{ e_{\beta}(X) - \mu + |c_{x_-}|^2 \Re\Sigma_{x_-}(i\omega_0) + |c_{z_-}|^2 \Re\Sigma_{z_-}(i\omega_0) \} \\ & \times \{ |c_{x_-}|^2 \sqrt{Z}_{x_-} + |c_{z_-}|^2 \sqrt{Z}_{z-} \},
\end{split}
\end{align}
where $\sqrt{Z}_{l}$ and $\Re\Sigma_{l}(i\omega_0)$ are the diagonal elements corresponding to the orbital $l$ of $\bm{\sqrt{Z}}$ and $\Re\bm{\Sigma}(i\omega_0)$, respectively. 
$|c_l|^2$ is the weight of the Wannier function $|\widetilde{d}_l \rangle$ in the $\beta$ pocket eigenstate $|\psi_{\beta}(\bm{k}) \rangle $ of $\bm{H_0}(\bm{k})$ at $\bm{k}=X$; $|\psi_{\beta}(\bm{k}=X) \rangle = \sum_{l}c_l  |\widetilde{d}_l \rangle$. We find that $c_{x_+} = c_{z_+} = 0$ in this case. 
Furthermore, near the $X$ point of the $\beta$ pocket $x_-$ and $z_-$ orbitals have almost equal weight as highlighted with dotted black ovals in Fig.~\ref{sfig_pressure}(c), so $|c_{x_-}|^2\simeq |c_{z_-}|^2 \simeq 0.5$. Thus, Eq.~(\ref{eqR8}) implies that the electronic correlations emerging from $x_-$ and $z_-$ orbital components influence $\hat{e}_{\beta}(X)$ as a weighted average.

Now, let us see the result of our analysis. Figure~\ref{sfig_pressure}(e) presents how $\hat{e}_{\beta}(X)$ changes with $t^z_\perp$ according to Eq.~(\ref{eqR7}). Indeed, $\hat{e}_{\beta}(X)$ increases with $|t^z_\perp|$ in CDMFT, which is consistent with how the $\beta$ pocket changes with $|t^z_\perp|$ in Fig.~\ref{sfig_pressure}(a). The reason for this increase of $\hat{e}_{\beta}(X)$ is because $\Re\Sigma_{z_-}(i\omega_0)$ is enhanced significantly with $t^z_\perp$ in CDMFT as presented in Fig.~\ref{sfig_pressure}(d), in contrast to the other terms of Eq.~(\ref{eqR7}) which are found to vary only weakly with $t^z_\perp$. Thus, how $\hat{e}_{\beta}(X)$ changes with pressure is almost entirely dependent on how $\Re\Sigma_{z_-}(i\omega_0)$ changes with $t^z_\perp$.
In this respect, it is useful to examine $\hat{e}_{\beta}(X)$ in DMFT since $\Re\Sigma_{z_-}(i\omega_0)$ weakly changes with $t^z_\perp$ in DMFT; see Fig.~\ref{sfig_pressure}(d). Indeed, $\hat{e}_{\beta}(X)$ remains almost unchanged with $t^z_\perp$ in DMFT as shown in Fig.~\ref{sfig_pressure}(e), which also explains why the $\beta$ pocket in DMFT remains almost intact.

\section{The low-pressure Fermi surface}

\begin{figure} [!htbp] 
	\includegraphics[width=0.9\textwidth]{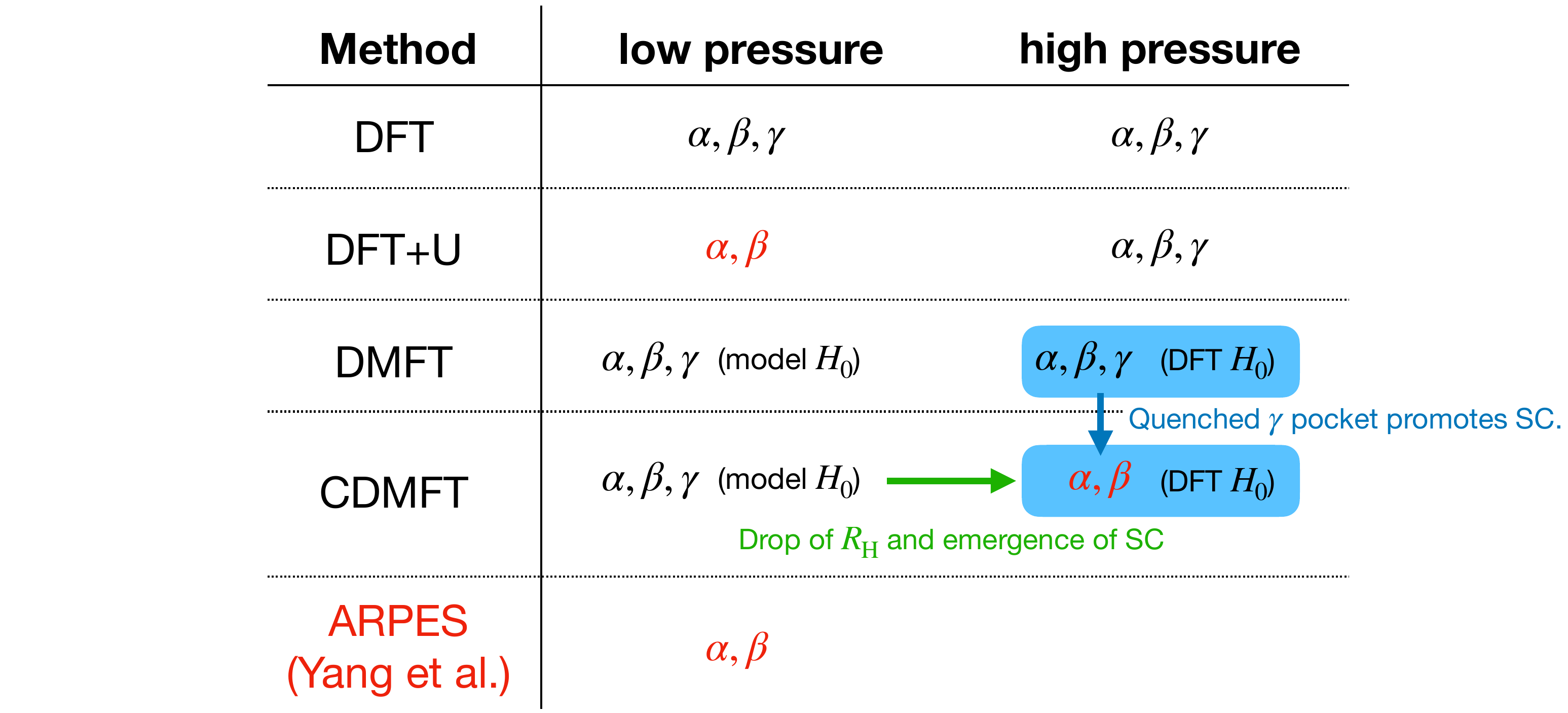}
	\caption{The table summarizes which method results in which FS pockets at the low-pressure and high-pressure phases of La$_3$Ni$_2$O$_7$.  Here ``model $H_0$" indicates the $H_0$ obtained from DFT in the high pressure phase with $t^z_\perp$ being tuned to mimic the low-pressure phase. The ARPES FS by Yang et al. \cite{yang_orbital-dependent_2023} for the ambient-pressure phase is composed of the $\alpha$ and the $\beta$ pockets. For the low-pressure DFT FS, see, e.g.,~Refs.~\cite{labollita_electronic_2023,zhang_structural_2023}. For the high-pressure DFT FS, see, e.g.,~Refs.~\cite{luo_bilayer_2023,gu_effective_2023,zhang_electronic_2023,cao_flat_2023,sakakibara_possible_2023,liu_spm-wave_2023,labollita_electronic_2023,zhang_structural_2023,zhang_trends_2023}. For both the low-pressure and the high-pressure DFT+$U$ FSs, see, e.g.,~Refs.~\cite{sun_signatures_2023,geisler_structural_2023}.
		Our high-pressure DMFT FS is also consistent with that in Refs.~\cite{shilenko_correlated_2023,lechermann_electronic_2023}.
	}
	\label{sfig_table}
\end{figure}

Figure~\ref{sfig_table} summarizes which theoretical method results in which FS pockets.  We once again clarify here that, to simulate the high-pressure phase, we have employed DMFT and CDMFT for the {\it ab initio} model  with one-body part of the Hamiltonian being derived from DFT (dubbed ``DFT $H_0$"), as highlighted in Fig.~\ref{sfig_table}.

Having established the role of the interlayer correlations for the high-pressure phase, it is tempting to reach out to a picture also commenting on the low-pressure phase. To this end, we have used a simplified model, where we have only rescaled $t^z_\perp$ of DFT $H_0$ (dubbed ``model $H_0$" in Fig.~\ref{sfig_table}). With this model we could possibly explain the evolution of $R_\mathrm{H}$  \cite{zhou2023evidence} and the emergence of superconductivity by comparing high- and low-pressure phases; see Fig.~\ref{sfig_table}. Furthermore, the calculated CDMFT spin susceptibility for the low-pressure ``model" is close to the spin-density wave (SDW) ordering vector reported by experiments  \cite{chen2024electronic, dan2024spindensitywave} as discussed in the main text and Sec.~\ref{sec:sus_sp}.

The $\gamma$ pocket is absent in the available ARPES data \cite{yang_orbital-dependent_2023}, which is not consistent with our ``low-pressure" phase results; see also Fig.~\ref{sfig_table}. Drawing a definite conclusion would be premature at the moment, however, considering the delicacy of ARPES experiment on nickelates due to the oxygen vacancies (which are prevalent in the actual La$_3$Ni$_2$O$_7$ samples), the sample-to-sample variation of $\mathrm{NiO}_6$-plane stacking in La$_3$Ni$_2$O$_7$ (trilayer--monolayer vs.~bilayer--bilayer), and surface effects. In this regard, it is noteworthy that there is a discrepancy exactly on the fermiology even between the two experiments on La$_4$Ni$_3$O$_{10}$. Namely, while Du et al.~\cite{du2024correlated} reported that the $\gamma$ pocket constitutes the FS and its position remains unchanged with temperature (presumably down to $\sim 20$~K), a previous ARPES study by Li et al.~\cite{dessau2017} reported in contrast that this $\gamma$ pocket is gapped out at a low temperature  ($\sim 30$~K), thereby being disappeared from the FS. Further theoretical as well as experimental study is requested to establish the fermiology of the ambient-pressure phase.

\section{How nesting of the $\beta$ Fermi-surface pocket is affected by interlayer correlations}

\begin{figure} [!htbp] 
	\includegraphics[width=1.0\textwidth]{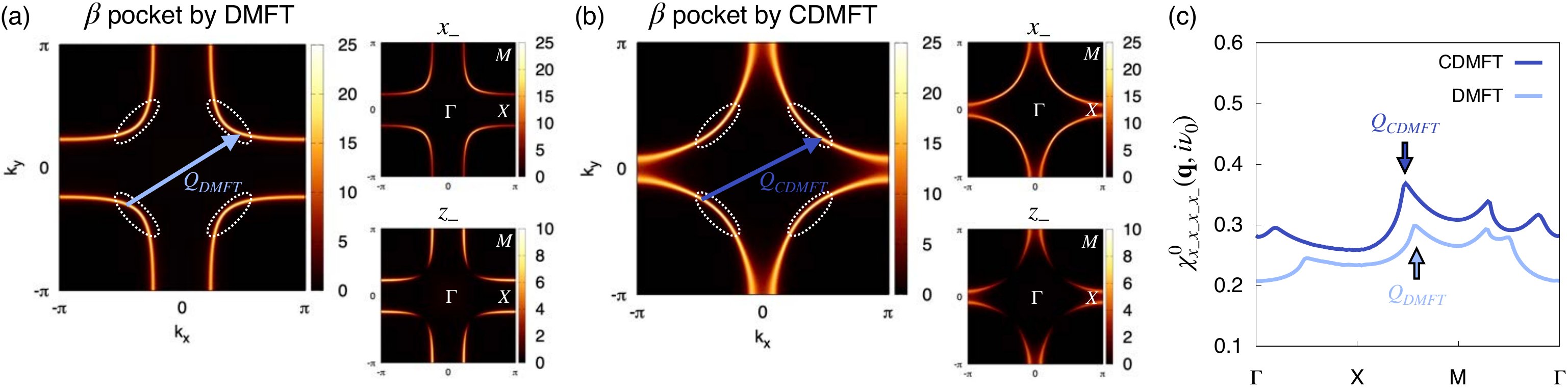}
	\caption{The $\beta$ Fermi surface pockets obtained from (a) DMFT and (b) CDMFT. The $\beta$ pocket consists of both $x_-$ and $z_-$ orbital character as presented in the right panels.  The nesting vector $\bm{Q}$ is highlighted by $\bm{Q}_{\mathrm{(C)DMFT}}$ . (c) $\chi^0_{x_- x_- x_- x_-}(\bm{q}, i\nu_0)$ calculated using DMFT and CDMFT Green's functions. $\bm{Q}_{\mathrm{(C)DMFT}}$ and the associated values of $\chi^0_{x_- x_- x_- x_-}(\bm{q}, i\nu_0)$ are highlighted with colored arrows.}
	\label{sfig_nesting}
\end{figure}

The $\beta$ Fermi surface pocket is also modified by the interlayer electronic correlations (IECs) in CDMFT as discussed in the previous section. This change is beneficial for the nesting of the irreducible susceptibility $\chi^0_{x_- x_- x_- x_-}(\bm{q}, i\nu_0)$ as shown in Fig.~\ref{sfig_nesting}.

In Fig.~\ref{sfig_nesting}(a) and Fig.~\ref{sfig_nesting}(b), we present the DMFT and CDMFT $\beta$ pockets and their associated orbital character. Since the shape of the $\beta$ pocket becomes more ``diamond-shaped" in CDMFT, the FS sectors in which the $x_-$ orbital occupies the entire weight, as highlighted with dotted white ovals in Fig.~\ref{sfig_nesting}(a) and Fig.~\ref{sfig_nesting}(b), become less curved. In addition to the shape change, spectral weight of $x_-$ orbital near the $X$ point gets enhanced in CDMFT. These changes in CDMFT result in i) an overall enhancement of  $\chi^0_{x_- x_- x_- x_-}(\bm{q}, i\nu_0)$ and ii) a better-nesting for the nesting vectors connecting the two dotted white ovals facing each other along the zone diagonal direction; see, e.g., $\bm{Q}_{\mathrm{DMFT}}$ and $\bm{Q}_{\mathrm{CDMFT}}$, as can be confirmed from Fig.~\ref{sfig_nesting}(c).

\section{Gap equation}

\subsection{Pairing interactions}

\begin{figure} [!htbp] 
	\includegraphics[width=0.4\textwidth]{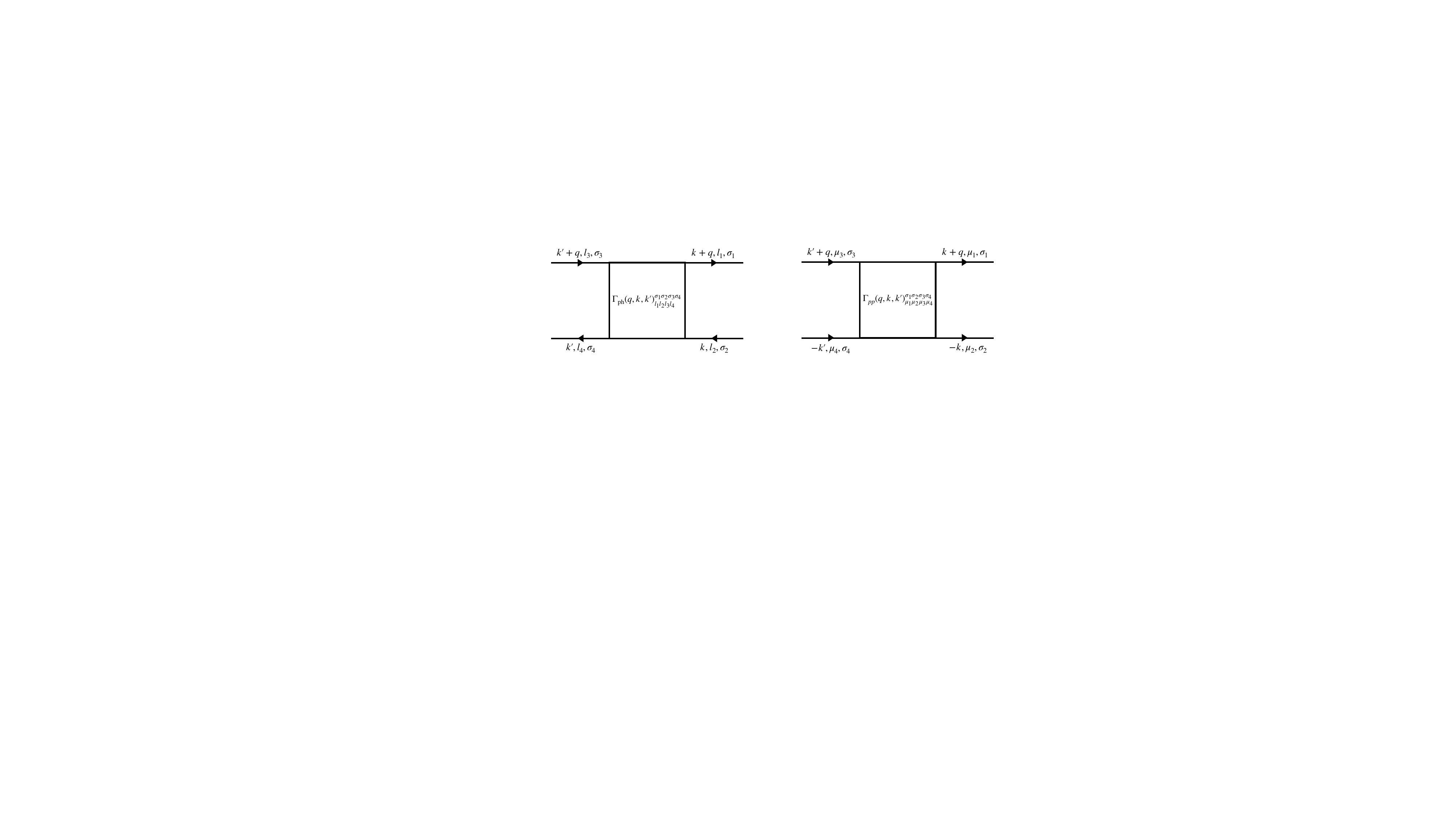}
	\caption{The particle-hole irreducible vertex.}
	\label{sfig_vertices}
\end{figure}

To obtain the effective singlet and triplet pairing interactions, we begin with the spin and charge irreducible vertices in the particle-hole channel, which read
\begin{align}
	\Gamma^{\mathrm{sp/ch}}(q,k,k')_{l_1 l_2 l_3 l_4}  = \Gamma_{\mathrm{ph}}(q,k,k')_{l_1 l_2 l_3 l_4}^{\uparrow \uparrow \downarrow \downarrow} \mp \Gamma_{\mathrm{ph}}(q,k,k')_{l_1 l_2 l_3 l_4}^{\uparrow \uparrow \uparrow \uparrow}.
\end{align}
$\Gamma_{\mathrm{ph}}(q,k,k')^{\sigma_1 \sigma_2 \sigma_3 \sigma_4 }_{l_1 l_2 l_3 l_4}$ is displayed in Fig.~\ref{sfig_vertices}. $k \equiv (\bm{k},i\omega_n)$ and $q \equiv (\bm{q},i\nu_n)$. 
The spin and charge susceptibilities are obtained using these vertices from the Bethe-Salpeter equation:
\begin{align}
	\chi^{\mathrm{sp/ch}}(q)^{l_1  l_2 l_7  l_8}_{kk'}  = \chi^{0}(q)^{l_1  l_2 l_7  l_8}_{kk'} \pm \frac{T^2}{N^2} \chi^{\mathrm{sp/ch}}(q)^{l_1  l_2 l_3  l_4}_{kk_1} \Gamma^{\mathrm{sp/ch}}(q,k_1,k_2)_{l_3 l_4 l_5 l_6} \chi^{0}(q)^{l_5  l_6 l_7  l_8}_{k_2 k'}, 
	\label{chi}
\end{align} 
where $\chi^{0}(q)^{l m l' m'}_{kk'} = -\frac{N}{T}G_{l l'}(k+q)G_{m' m}(k')\delta_{kk'}$. Here indices repeated twice should be summed over. We below keep this summation convention.
By employing the parquet equations, one can formally express the singlet (s) and triplet (t) pairing interactions using $\bm{\Gamma}^{\mathrm{sp/ch}}$ and $\bm{\chi}^{\mathrm{sp/ch}}$ as \cite{bickers_self-consistent_2004,rohringer_diagrammatic_2018}
\begin{align}
\begin{split}
	\Gamma^{\mathrm{s}}_{l_1 l_4 l_3 l_2}(k,k') = \Lambda^{\mathrm{irr,s}}(k,k')_{l_1 l_4 l_3 l_2} &+ \frac{3}{2} \Phi^{\mathrm{sp}}(-k+k',-k',k)_{l_2 l_4 l_3 l_1}  -   \frac{1}{2} \Phi^{\mathrm{ch}}(-k+k',-k',k)_{l_2 l_4 l_3 l_1} \\
	&+ \frac{3}{2} \Phi^{\mathrm{sp}}(k+k',-k',-k)_{l_1 l_4 l_3 l_2 }  -   \frac{1}{2} \Phi^{\mathrm{ch}}(k+k',-k',-k)_{l_1 l_4 l_3 l_2 }, 
\end{split} \label{eqS8} \\
\begin{split}
	\Gamma^{t}_{l_1 l_4 l_3 l_2}(k,k') = \Lambda^{\mathrm{irr,t}}(k,k')_{l_1 l_4 l_3 l_2} &- \frac{1}{2} \Phi^{\mathrm{sp}}(-k+k',-k',k)_{l_2 l_4 l_3 l_1}  -   \frac{1}{2} \Phi^{\mathrm{ch}}(-k+k',-k',k)_{l_2 l_4 l_3 l_1} \\
	&- \frac{1}{2} \Phi^{\mathrm{sp}}(k+k',-k',-k)_{l_1 l_4 l_3 l_2 }  -   \frac{1}{2} \Phi^{\mathrm{ch}}(k+k',-k',-k)_{l_1 l_4 l_3 l_2 },
\end{split} \label{eqS9} 
\end{align}
where ladder vertices $\Phi^{\mathrm{sp/ch}}(q,k,k')_{l_1 l_2 l_7 l_8} = \frac{T^2}{N^2} \Gamma^{\mathrm{sp/ch}}(q,k,k_1)_{l_1 l_2 l_3 l_4} \chi^{\mathrm{sp/ch}}(q)^{l_3 l_4 l_5  l_6}_{k_1 k_2}  \Gamma^{\mathrm{sp/ch}}(q,k_2,k')_{l_5 l_6 l_7 l_8}$. $\Lambda^{\mathrm{irr,s/t}}(k,k')_{l_1 l_4 l_3 l_2}$ are the fully irreducible vertices in the singlet (s) and triplet (t) channels.  The bare constant terms of $\bm{\Lambda}^{\mathrm{irr,s/t}}$, namely $\bm{\Lambda}^{0,\mathrm{s/t}}$, are linear combinations of bare spin and charge interaction tensors $\bm{\mathcal{U}}^{\mathrm{sp/ch}}$:
 $\Lambda^{0,\mathrm{s}}_{l_1 l_4 l_3 l_2} = \frac{1}{2}[3\mathcal{U}^{\mathrm{sp}}_{} + \mathcal{U}^{\mathrm{ch}}]_{l_1 l_4 l_3 l_2}$ and $\Lambda^{0,\mathrm{t}}_{l_1 l_4 l_3 l_2} = -\frac{1}{2}[\mathcal{U}^{\mathrm{sp}} -\mathcal{U}^{\mathrm{ch}}]_{l_1 l_4 l_3 l_2}$.  Since the interaction tensor $H_\mathrm{int}$ [Eq.~(\ref{Hint})] is more sparse in the original $e_g$-orbital basis than the BA basis, so are the resulting $\bm{\mathcal{U}}^{\mathrm{sp/ch}}$ which are given by $\mathcal{U}^{\mathrm{sp/ch}}_{l_1 l_4 l_3 l_2} = \mathcal{U}^{\uparrow \downarrow}_{l_1 l_3 l_4 l_2} \mp (\mathcal{U}^{\uparrow \uparrow}_{l_1 l_3 l_4 l_2} - \mathcal{U}^{\uparrow \uparrow}_{l_1 l_4 l_3 l_2})$. In the $e_g$-orbital basis, elements of  $\bm{\mathcal{U}}^{\mathrm{sp/ch}}$  read
\begin{align}
	\mathcal{U}^{\mathrm{sp}}_{l_1 l_4 l_3 l_2} = \left\{
	\begin{aligned}
		&U\\ &U' \\ &J \\ &J
	\end{aligned}\right.\;,\quad
	\mathcal{U}^{\mathrm{ch}}_{l_1 l_4 l_3 l_2} = \left\{
	\begin{aligned}
		&U &\mathrm{if}\, l_1= l_2 = l_3 = l_4,\\
		&-U' + 2J &\mathrm{if}\, l_1 = l_3 \neq l_2 = l_4, \\
		&2U' -J &\mathrm{if}\, l_1 = l_4 \neq l_2 = l_3, \\
		&J&\mathrm{if}\, l_1 = l_2 \neq l_3 = l_4,
	\end{aligned}\right.
\end{align}
for the Kanamori interaction with $U'=U-2J$.
Using above formulas, a superconducting instability can be formulated in terms of a non-Hermitian eigenvalue problem, namely the gap equation, which reads
\begin{align}
	\lambda_\mathrm{sc} \Delta_{l_1 l_2}(k) =& -\frac{T}{2N} \sum_{k', l_3 l_4 l_5 l_6 } \Gamma^{\mathrm{s/t}}_{l_1  l_4 l_3 l_2}(k,k')G_{l_4 l_5} (k') G_{l_3 l_6}(-k') \Delta_{l_5 l_6}(k'),
\end{align} 
where $\lambda_\mathrm{sc}$ is the eigenvalue, $G_{l m}(k)$ the (C)DMFT Green's function, and $\Delta_{l m}(k)$ the anomalous self-energy (gap function). The transition temperature $T_{\mathrm{c}}$ corresponds to the temperature at which the maximum (leading) eigenvalue $\lambda_\mathrm{sc}$ reaches unity.

\subsection{Approximations for the pairing interactions and the resulting gap equation}
Evaluating Eqs.~(\ref{eqS8}) and (\ref{eqS9}) for the full vertex is a formidable task, it thus requires some approximations. First, we employ the well-known parquet approximation, which approximates $\bm{\Lambda}^{\mathrm{irr,s/t}}$ by $\bm{\Lambda}^{0,\mathrm{s/t}}$. Within DMFT and CDMFT for our model, $\Gamma^{\mathrm{sp/ch}}(q,k,k')_{l_1 l_2 l_3 l_4}$ can be approximated by the momentum-independent impurity vertices $\Gamma^{\mathrm{sp/ch}}_{\mathrm{imp}}(i\nu_m,i\omega_n,i\omega_{n'})_{l_1 l_2 l_3 l_4}$, which are in principle feasible to be numerically evaluated using the continuous-time quantum Monte Carlo methods \cite{CTQMC}. However, for multiorbital systems huge stochastic noise prohibits us from using the measured $\Gamma^{\mathrm{sp/ch}}_{\mathrm{imp}}(i\nu_m,i\omega_n,i\omega_{n'})_{l_1 l_2 l_3 l_4}$ in practice. Following an idea employed in Refs.~\cite{maier_spin_2007,nourafkan_correlation-enhanced_2016,gingras_superconducting_2019,kaser_interorbital_2022}, we parametrize $\bm{\Gamma}^{\mathrm{sp/ch}}$ using effective screened Coulomb interaction $\overline{U}$ and Hund's coupling $\overline{J}$, i.e., we substitute $\bm{\Gamma}^{\mathrm{sp/ch}}  \rightarrow \overline{\bm{\Gamma}}^{\mathrm{sp/ch}}(\overline{U},\overline{J})$. As we used for the bare interaction, the effective screened interorbital Coulomb interaction  $\overline{U}'$ obeys $\overline{U}'= \overline{U}-2\overline{J}$.  Note that $\overline{U}$ and $\overline{J}$ are generally different from bare $U$ and $J$ which enter $\bm{\Lambda}^{0,s/t}$. In the $e_g$-orbital basis, $\overline{\bm{\Gamma}}^{\mathrm{sp/ch}}(\overline{U},\overline{J})$ are then given by
\begin{align}
	\overline{\Gamma}^{\mathrm{sp}}_{l_1 l_4 l_3 l_2} = \left\{
	\begin{aligned}
		&\overline{U}\\ &\overline{U}' \\ &\overline{J} \\ &\overline{J}
	\end{aligned}\right. \;,\quad
	\overline{\Gamma}^{\mathrm{ch}}_{l_1 l_4 l_3 l_2} = \left\{
	\begin{aligned}
		&\overline{U} &\mathrm{if}\, l_1= l_2 = l_3 = l_4,\\
		&-\overline{U}' + 2\overline{J} &\mathrm{if}\, l_1 = l_3 \neq l_2 = l_4, \\
		&2\overline{U}' -\overline{J} &\mathrm{if}\, l_1 = l_4 \neq l_2 = l_3, \\
		&\overline{J}&\mathrm{if}\, l_1 = l_2 \neq l_3 = l_4,
	\end{aligned}\right.
 \label{eq:vertex_local_static}
\end{align}
Using this approximation,
\begin{align}
\Phi^{\mathrm{sp/ch}}(q,k,k')_{l_1 l_2 l_7 l_8}  \rightarrow \overline{\Phi}^{\mathrm{sp/ch}}(q)_{l_1 l_2 l_7 l_8} = \overline{\Gamma}^{\mathrm{sp/ch}}_{l_1 l_2 l_3 l_4} \chi^{\mathrm{sp/ch}}(q)_{l_3 l_4 l_5 l_6} \overline{\Gamma}^{\mathrm{sp/ch}}_{l_5 l_6 l_7 l_8}.
\end{align}
Here $ \chi^{\mathrm{sp/ch}}(q)_{l_1 l_2 l_3 l_4} \equiv \frac{T^2}{N^2}\sum_{kk'} \chi^{\mathrm{sp/ch}}(q)^{l_1 l_2 l_3 l_4}_{kk'}$ obtained from using $\overline{\bm{\Gamma}}^{\mathrm{sp/ch}}(\overline{U},\overline{J})$ for the irreducible vertices in Eq.~(\ref{chi}):
\begin{align}
\begin{split}
 \chi^{\mathrm{sp/ch}}(q)_{l_1 l_2 l_3 l_4} \equiv \frac{T^2}{N^2}\sum_{kk'} \chi^{\mathrm{sp/ch}}(q)^{l_1 l_2 l_3 l_4}_{kk'}  &= \chi^{0}(q)_{l_1  l_2 l_7  l_8} \pm \chi^{\mathrm{sp/ch}}(q)_{l_1  l_2 l_3  l_4} \overline{\Gamma}^{\mathrm{sp/ch}}_{l_3 l_4 l_5 l_6} \chi^{0}(q)_{l_5  l_6 l_7  l_8}, \\
 &= \big[ \bm{\chi}^{0}(q)[\mathds{1} \mp \bm{\overline{\Gamma}}^{\mathrm{sp/ch}} \bm{\chi}^{0}(q)]^{-1} \big]_{l_1 l_2 l_3 l_4},
 \label{eq:RPA_sp_ch_susceptibility}
\end{split}
\end{align}
where $\chi^{0}(q)_{l_1  l_2 l_3 l_4} \equiv \frac{T^2}{N^2}\sum_{kk'} \chi^{0}(q)^{l_1  l_2 l_3 l_4}_{kk'}  = - \frac{T}{N} \sum_{k}G_{l_1 l_3}(k+q)G_{l_4 l_2}(k) $ and $\mathds{1}$ is identity matrix in orbital space.
We finally arrive at the gap equation presented in the main text
\begin{align}
	\begin{split}
		\lambda_\mathrm{sc} \Delta_{l_1 l_2}(k) =& -\frac{T}{2N} \sum_{k', l_3 l_4 l_5 l_6 } \Gamma^{\mathrm{s/t}}_{l_1  l_4 l_3 l_2}(k-k') G_{l_4 l_5} (k') G_{l_3 l_6}(-k') \Delta_{l_5 l_6}(k').
	\end{split}
	\label{gapeq}
\end{align}
Using the symmetry relations of singlet and triplet gap functions \cite{gingras_superconducting_2019,kaser_interorbital_2022} the effective singlet and triplet pairing interactions $\bm{\Gamma}^{\mathrm{s/t}}$ in Eq.~(\ref{gapeq}) are given by
\begin{align}
\begin{split}
\Gamma^{\mathrm{s}}(k-k')_{l_1 l_4 l_3 l_2} &= \Lambda^{0,\mathrm{s}}_{l_1 l_4 l_3 l_2}  + 3\overline{\Phi}^{\mathrm{sp}}(k-k')_{l_1 l_4 l_3 l_2} - \overline{\Phi}^{\mathrm{ch}}(k-k')_{l_1 l_4 l_3 l_2}, \\
&= \Lambda^{0,\mathrm{s}}_{l_1 l_4 l_3 l_2}  + 3[\overline{\bm{\Gamma}}^\mathrm{sp} \bm{\chi}^\mathrm{sp}(k-k') \overline{\bm{\Gamma}}^\mathrm{sp}]_{l_1 l_4 l_3 l_2} - [\overline{\bm{\Gamma}}^\mathrm{ch} \bm{\chi}^\mathrm{ch}(k-k') \overline{\bm{\Gamma}}^\mathrm{ch}]_{l_1 l_4 l_3 l_2},
\end{split}\label{Gamma_s}  \\
\begin{split}
\Gamma^{\mathrm{t}}(k-k')_{l_1 l_4 l_3 l_2} &= \Lambda^{0,\mathrm{t}}_{l_1 l_4 l_3 l_2}  - \overline{\Phi}^{\mathrm{sp}}(k-k')_{l_1 l_4 l_3 l_2} - \overline{\Phi}^{\mathrm{ch}}(k-k')_{l_1 l_4 l_3 l_2}, \\
&= \Lambda^{0,\mathrm{t}}_{l_1 l_4 l_3 l_2}  - [\overline{\bm{\Gamma}}^\mathrm{sp} \bm{\chi}^\mathrm{sp}(k-k') \overline{\bm{\Gamma}}^\mathrm{sp}]_{l_1 l_4 l_3 l_2} - [\overline{\bm{\Gamma}}^\mathrm{ch} \bm{\chi}^\mathrm{ch}(k-k') \overline{\bm{\Gamma}}^\mathrm{ch}]_{l_1 l_4 l_3 l_2}.
\end{split}
\end{align}

\subsection{Solving the gap equation using the intermediate representation basis}
\begin{figure} [!t] 
    \includegraphics[width=1.0\textwidth]{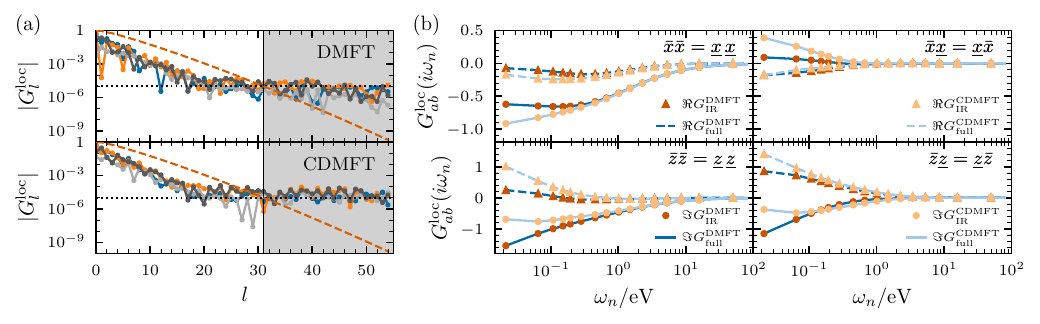}
    \caption{Representation of local (C)DMFT Green's functions $G^{\mathrm{loc}}_l = \frac{1}{N_{\bm{k}}}\sum_{\bm{k}} G_l(\bm{k})$ in the IR basis. (a) Expansion coefficients $G_l$ as a function of expansion order $l$. Colored lines with dots correspond to coefficients for different orbital components in the $e_g$ basis. The red dashed line corresponds to the truncation error set by the singular values. We cut the expansion after $l_{\rm max} = 31$ coefficients (vertical solid line) with a truncation error of $\delta_{\mathrm{IR}} = 10^{-5}$ (horizontal dotted line), such that the coefficients in the gray shaded area are disregarded. (b) Reconstruction of the (C)DMFT Green's function orbital components ($a,b\in\lbrace\bar{x},\bar{z},\underline{x},\underline{z}\rbrace$) on a sparse frequency grid from the IR basis. Markers and lines represent data on the sparse IR grid and full equidistant grid, respectively. Note the logarithmic frequency axis.}
    \label{sfig_IR_expansion}
\end{figure}

We solve the gap equation [Eq.~(\ref{gapeq})] on a $96\times 96$ $\bm{k}$-mesh using the power iteration method. We converge up to a tolerance of $5\cdot 10^{-5}$ for the eigenvalue $\lambda_{\mathrm{sc}}$ of the dominant gap function. In order to efficiently solve this self-consistent problem, we use the sparse-sampling approach~\cite{li_sparse_ir_2020,witt_efficient_2021,witt_twisted_tmdc_2022,witt_lk99_2023} of the intermediate representation (IR) basis~\cite{shinaoka_irbasis_2017,shinaoka_lecture_irbasis_2022} and its python library \texttt{sparse-ir}~\cite{wallerberger_sparse-ir_2023} for efficient data compression. To set up the IR basis functions, we use the IR basis parameter $\Lambda_{\mathrm{IR}} = \beta\omega_{\mathrm{max}} = 1450$ with inverse temperature $\beta = 145$\,eV$^{-1}\simeq 80$\,K and cutoff frequency $\omega_{\mathrm{max}} = 10$\,eV. We transform the (C)DMFT lattice Green's function $G(\bm{k}, i\omega_n) = \left[(i\omega_n + \mu)\mathds{1} - H_0(\bm{k}) - \bm{\Sigma}(i\omega_n)\right]^{-1}$ data ($\mathds{1}$ is unity matrix in orbital space and $\mu$ the chemical potential) on its full and equidistantly sampled frequency mesh ($N_{\omega_n} = 1846$ for $\omega_n>0$) to the IR basis functions $U_l$ from the expansion
\begin{align}
    G(i\omega_n) = \sum_{l=0}^{l_{\mathrm{max}}} G_l U_l(i\omega_n)
\end{align}
via least-square fitting. We set the truncation error to $\delta_{\mathrm{IR}} = 10^{-5}$ to not overfit the statistical noise from the quantum Monte Carlo (QMC) simulations. This corresponds to using $N_{\mathrm{IR}} = l_{\mathrm{max}} - 1 = 31$ basis coefficients. Note, that we here use the notation $ab$ instead of $lm$ for orbital indices $\lbrace\bar{x},\bar{z},\underline{x},\underline{z}\rbrace$ in the $e_g$ basis to prevent confusion with the IR basis index $l$. We set the basis size from analyzing the decay of the local (C)DMFT Green's function expansion coefficients $G^{\mathrm{loc}}_l$ [Fig.~\ref{sfig_IR_expansion}(a)]. Beyond $l_{\mathrm{max}}$ (gray shaded area), the coefficients do not decay anymore and become larger than the exponentially decaying singular values $S_l/S_0$ (red dashed line) of the IR basis kernel due to fitting of QMC noise. The singular values approximately set the truncation error~\cite{wallerberger_sparse-ir_2023}. By taking coefficients larger than the singular values into account, the conditioning of the transformation between IR basis and Matsubara frequency/imaginary time becomes bad, i.e., the numerical error can potentially amplify in an uncontrolled manner during calculation. Fig.~\ref{sfig_IR_expansion}(b) shows the corresponding reconstructed Green function data on a sparsely sampled frequency grid with $N_{\omega_n}^{\mathrm{IR}} = l_{\mathrm{max}} = 32$ frequency points.

\section{Mapping of interaction values to Stoner enhancement factors}\label{sec:Stoner_factors}
Throughout the paper, we discuss the superconducting phase diagram in terms of the Stoner enhancement factor $\alpha_{\mathrm{sp}}$. It is a proxy for the system's tendency towards a magnetic instability that we obtain from analyzing where the spin susceptibility $\chi^{\mathrm{sp}}$ [Eq.~(\ref{eq:RPA_sp_ch_susceptibility})] diverges, i.e., where the denominator becomes zero. This is true, if the largest eigenvalue $\alpha_{\mathrm{sp}} = \mathrm{max}_q\lbrace \bm{\overline{\Gamma}}^{\mathrm{sp}}\bm{\chi}^0(q)\rbrace$ reaches unity. Here, we comment further on the Stoner enhancement of the (C)DMFT calculations.

\begin{figure} [!t] 
    \includegraphics[width=1\textwidth]{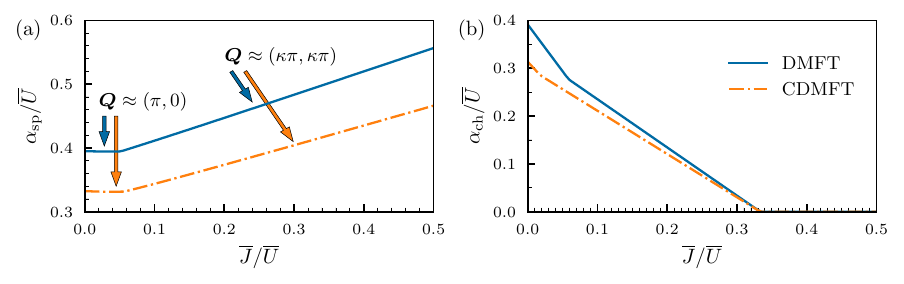}
    \caption{Stoner enhancement factor in the (a) magnetic and (b) charge channel for the irreducible susceptibility of the (C)DMFT as a function of interaction ratio $\overline{J}/\overline{U}$. The kink around $\overline{J}/\overline{U}\approx 0.05$\,--\,0.06 originates from different $\bm{Q}$ vectors contributing to the maximal eigenvalue of  $\overline{\bm{\Gamma}}^{\mathrm{sp}}\bm{\chi}^0(\bm{Q},i\nu_0)$ as indicated by arrows.}
    \label{sfig_Stoner_UJ_plane}
\end{figure}

For a given $\bm{\chi}^0(q)$, $\alpha_{\mathrm{sp}}$ is a a function of the effective Hund's coupling $\overline{J}$ and the effective intraorbital Coulomb interaction $\overline{U}$. By fixing the value of $\overline{J}/\overline{U}$, the ratio $\alpha_{\mathrm{sp}}/\overline{U}$ is uniquely determined, as we show in Fig.~\ref{sfig_Stoner_UJ_plane}(a). This function exhibits a kink around $\overline{J}/\overline{U}\approx 0.05$\,--\,0.06 which is manifested as a kink in the superconducting phase boundary ($\lambda_{\mathrm{sc}} = 1$) shown in Fig.~3(a) of the main text and Fig.~(\ref{sfig_scaling}). The kink originates from two different wave vectors $\bm{Q}$ at which the eigenvalue of $\overline{\bm{\Gamma}}^{\mathrm{sp}}\bm{\chi}^0(\bm{Q},i\nu_0)$ has its maximal value: around the X point $\bm{Q}\approx(\pi,0)$ for $\overline{J}/\overline{U}\lessapprox 0.06$ and an incommensurate wave vector around the M point for $\overline{J}/\overline{U}\gtrapprox 0.06$ which we here denote by $\bm{Q}\approx(\kappa\pi,\kappa\pi)$ with $\kappa< 1$. The change between these two dominant momenta $\bm{Q}$ depends on the momentum structure of $\bm{\chi}^0(q)$ [c.f.~Fig.~3(b) of the main text and Fig.~\ref{sfig_gap_vertex_structure} in Sec.~\ref{sec:gap_vertex}] and on how $\overline{J},\,\overline{U}$ mix different orbital components of $\bm{\chi}^0(q)$. The exact location of the kink is, hence, different for the DMFT and CDMFT irreducible susceptibilities. It is noticeable that $\alpha_{\mathrm{sp}}/\overline{U}$ of the DMFT calculation is generally larger than that of the CDMFT calculation. 

For completeness, we mention that an identical analysis of possible charge instabilities can be made. Here, the charge Stoner enhancement $\alpha_{\mathrm{ch}} = \mathrm{max}_q\lbrace -\bm{\overline{\Gamma}}^{\mathrm{ch}}\bm{\chi}^0(q)\rbrace$ needs to be analyzed [Fig.~\ref{sfig_Stoner_UJ_plane}(b)]. This system, however, does not host any charge instability because $\alpha_{\mathrm{ch}}$ is too small.

\section{Spin susceptibility}\label{sec:sus_sp}

\begin{figure} [!htbp] 
	\includegraphics[width=0.8\textwidth]{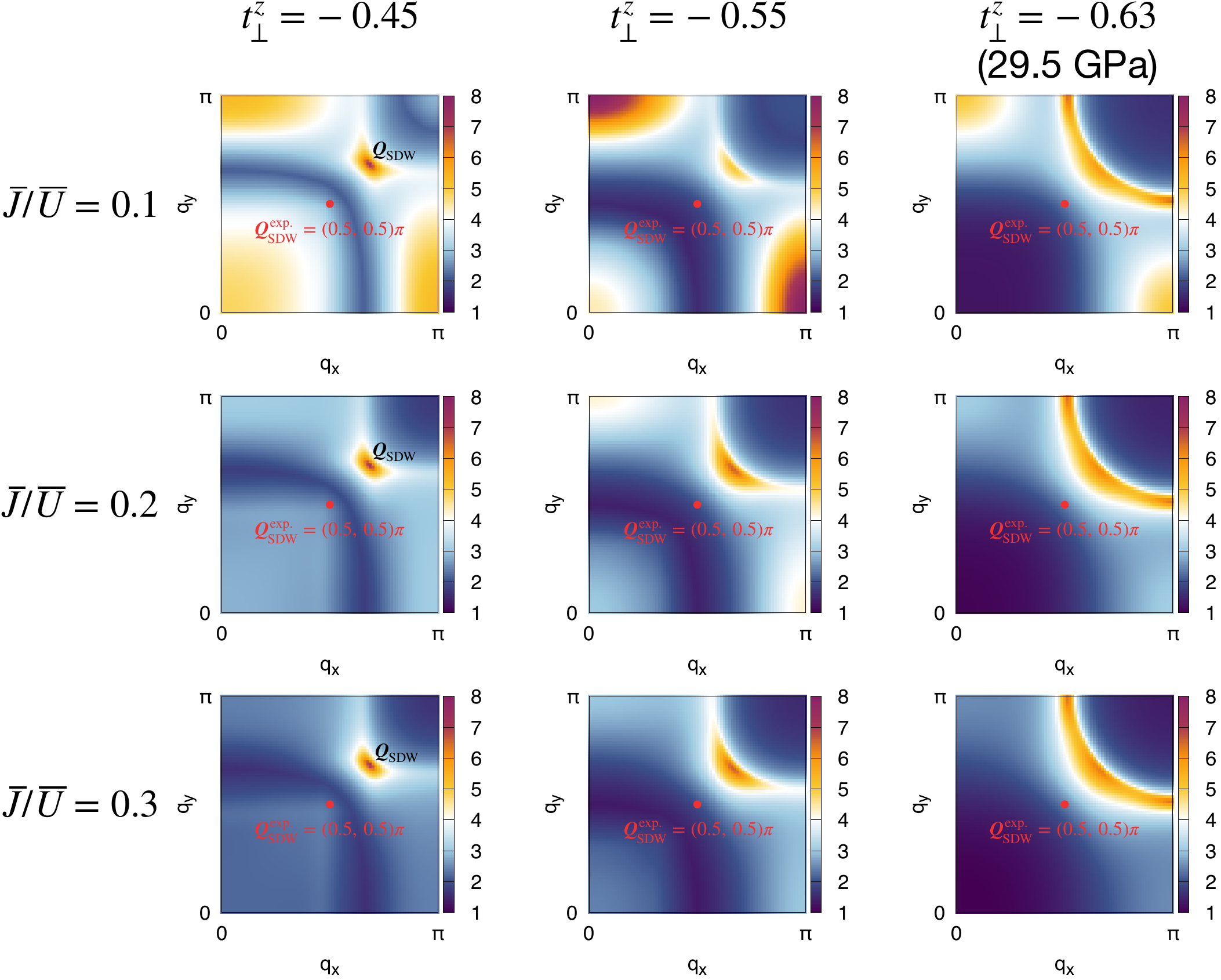}
	\caption{$\chi^\mathrm{sp}_\mathrm{max}(\bm{q}, i\nu_0)$ at $\alpha_\mathrm{sp}=0.95$. The suggested magnetic ordering vector under ambient pressure in experiments, $\bm{Q}^\mathrm{exp}_\mathrm{SDW}$  \cite{chen2024electronic, dan2024spindensitywave}, is highlighted with red circles in each panel. }
	\label{sfig_chi_max}
\end{figure}

While the $\gamma$ pocket gives rise to strong ferromagnetic fluctuations, the actual magnetic transition occurs at a finite transfer momentum as shown in Fig.~\ref{sfig_chi_max}. Figure~\ref{sfig_chi_max} presents the calculated maximum eigenvalue of the spin susceptibility at each $\bm{q}$, $\chi^\mathrm{sp}_\mathrm{max}(\bm{q}, i\nu_0)$. A peak of $\chi^\mathrm{sp}_\mathrm{max}(\bm{q}, i\nu_0)$ at $\bm{q}=\bm{Q}$ would result in a spin-density wave (SDW) with the ordering vector  $\bm{Q}$.

For $t^z_\perp = -0.45$~eV, $\chi^\mathrm{sp}_\mathrm{max}(\bm{q}, i\nu_0)$ shows a peak at $\bm{Q}_\mathrm{SDW}$ which is different from but close to the SDW ordering wave vector $\bm{Q}^\mathrm{exp}_\mathrm{SDW} = (0.5, 0.5)\pi$ (or $\bm{Q}^\mathrm{exp}_\mathrm{SDW} = (0.25, 0.25)$ if we follow the notation of Ref.~\cite{chen2024electronic}) reported by experiments under ambient pressure \cite{chen2024electronic, dan2024spindensitywave}.  On the other hand, in the high-pressure phase ($t^z_\perp = -0.63$~eV) $\chi^\mathrm{sp}_\mathrm{max}(\bm{q}, i\nu_0)$ forms an arc not a peak; see the rightmost panels in Fig.~\ref{sfig_chi_max}.  The strong ferromagnetic fluctuation by the $\gamma$ pocket is captured in Fig.~\ref{sfig_chi_max} in that depressurizing the system (by reducing the magnitude of $t^z_\perp$) enhances $\chi^\mathrm{sp}_\mathrm{max}(\bm{q}, i\nu_0)$ at $\bm{q}=0$, in accordance with the evolution of the Fermi surface in Fig.~4(b) of the main text.

\begin{figure} [!htbp] 
	\includegraphics[width=1.0\textwidth]{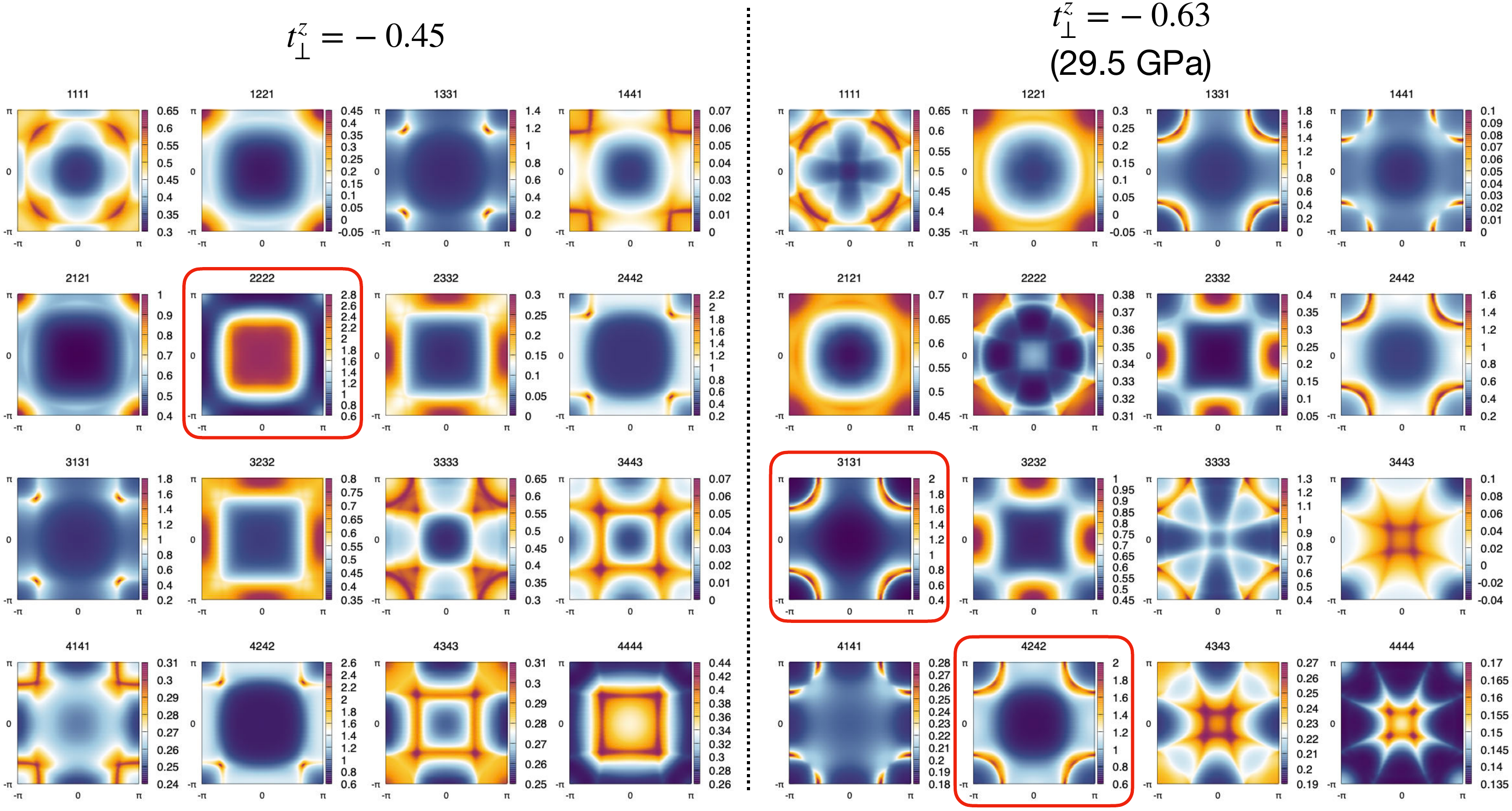}
	\caption{The orbital-resolved spin susceptibility $\chi^\mathrm{sp}_{lmml}(\bm{q}, i\nu_0)$ and $\chi^\mathrm{sp}_{lmlm}(\bm{q}, i\nu_0)$ at $\alpha_\mathrm{sp}=0.95$ and $\overline{J}/\overline{U}=0.2$. Indices 1, 2, 3, and 4 correspond to $x_+$, $z_+$, $x_-$, and $z_-$ orbitals, respectively. The matrix elements highlighted with red boxes correspond to the largest elements for each $t^z_\perp$.}
	\label{sfig_chi_matrix}
\end{figure}

Importantly, the aforementioned peak of $\chi^\mathrm{sp}_\mathrm{max}(\bm{q}, i\nu_0)$ at $\bm{q}=\bm{Q}_\mathrm{SDW}$ in the low-pressure phase is not driven by the $\gamma$ pocket, but by the $\alpha$--$\beta$ nesting. To see this, we present in Fig.~\ref{sfig_chi_matrix} the orbital-resolved spin susceptibilities, $\chi^\mathrm{sp}_{lmml}(\bm{q}, i\nu_0)$ and $\chi^\mathrm{sp}_{lmlm}(\bm{q}, i\nu_0)$ for both the smallest and the largest $t^z_\perp$ we used. For simplicity we hereafter use indices 1, 2, 3, and 4 for $x_+$, $z_+$, $x_-$, and $z_-$ orbitals, respectively. 
Interestingly, the left panel in Fig.~\ref{sfig_chi_matrix} shows that $\chi^\mathrm{sp}_{2222}(\bm{q}, i\nu_0)$ is the largest element for $t^z_\perp = -0.45$~eV and exhibits the maximum at $\bm{q}=0$ because of the $\gamma$ pocket crossing the Fermi level. It looks first at odds with $\chi^\mathrm{sp}_\mathrm{max}(\bm{q}, i\nu_0)$ presented in Fig.~\ref{sfig_chi_max}, where $\bm{q}$ shows a peak at $\bm{Q}_\mathrm{SDW}$, not at $\bm{q}=0$. It should be noted that the two subleading elements, namely $\chi^\mathrm{sp}_{3131}(\bm{q}, i\nu_0)$  and $\chi^\mathrm{sp}_{4242}(\bm{q}, i\nu_0)$ show the peak at $\bm{q} = \bm{Q}_\mathrm{SDW}$. Thus these two elements dominate over $\chi^\mathrm{sp}_{2222}(\bm{q}, i\nu_0)$, giving rise to the largest value of $\chi^\mathrm{sp}_\mathrm{max}(\bm{q}, i\nu_0)$ at $\bm{q} = \bm{Q}_\mathrm{SDW}$. The $\bm{Q}_\mathrm{SDW}$ corresponds to the vector connecting the $\alpha$ and the $\beta$ FS pockets as depicted in Fig.~4(b) of the main text. The discrepancy between the $\bm{Q}^\mathrm{exp}_\mathrm{SDW}$ and $\bm{Q}_\mathrm{SDW}$ can be attributed to the size of the $\alpha$ and $\beta$ pockets in simulating the low-pressure phases for which we neglected the change of in-plane hoppings due to the octahedral distortions present in the ambient-pressure structure.

In the high-pressure phase ($t^z_\perp = -0.63$~eV) $\chi^\mathrm{sp}_{2222}(\bm{q}, i\nu_0)$ is quenched due to the disappearance of the $\gamma$ pocket; see the right panel in Fig.~\ref{sfig_chi_matrix}. Here $\chi^\mathrm{sp}_{3131}(\bm{q}, i\nu_0)$  and $\chi^\mathrm{sp}_{4242}(\bm{q}, i\nu_0)$ are the largest elements which lead to the strong pair scattering between the bonding $\alpha$ and the antibonding $\beta$ pockets in line with the $s\pm$-wave gap symmetry.

\section{Additional data: Structure of pairing vertex and the $s\pm$-wave gap function}\label{sec:gap_vertex}
In the main text, we discuss the dominant pairing symmetry which is the intraorbital $s$-wave/interorbital $d_{x^2-y^2}$-wave pairing in the original $e_g$-orbital basis, which corresponds to the $s \pm$-wave symmetry by projecting it to the noninteracting FS. Apart from this symmetry, we checked many possible trial gap functions with different sign combinations of spin ($S$), parity ($P$), orbital ($O$), and time-reversal ($T$) symmetry which are in line with the $D_{4h}$ symmetry of the model. Importantly, throughout the whole $\alpha_{\mathrm{sp}}$\,-\,$\overline{J}/\overline{U}$ plane, none of these pairing channels have an eigenvalue larger than that of the $s \pm$-wave symmetry. Only a subleading channel with  intraorbital $d_{x^2-y^2}$/interorbital $s$-wave pairing reaches for the CDMFT electronic structure an eigenvalue $\lambda_{\mathrm{sc}}$ of unity in the region of $\alpha_{\mathrm{sp}}\gtrapprox0.96$ and $\overline{J}/\overline{U}\gtrapprox0.06$.

\begin{figure} [!htbp] 
    \includegraphics[width=1.0\textwidth]{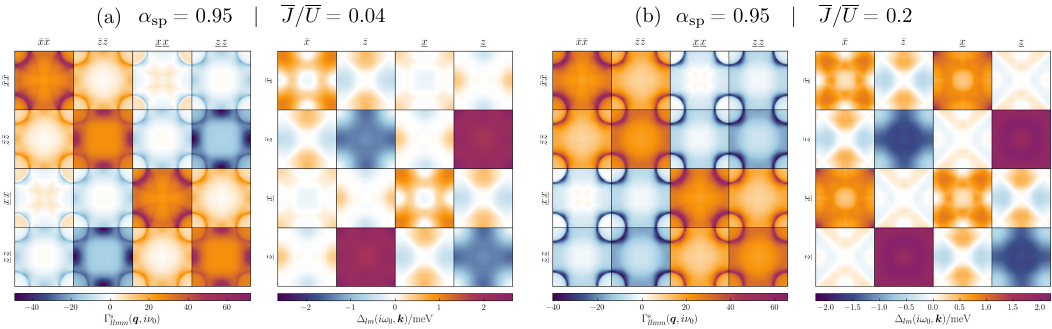}
    \caption{Orbital and momentum space structure of the dominant singlet pairing vertex components $\Gamma^{\mathrm{s}}_{llmm}(\bm{q},i\nu_0)$ and gap function $\Delta_{lm}(\bm{k},i\omega_0)$ for $\alpha_{\mathrm{sp}} = 0.95$ and two values of the interaction ratio $\overline{J}/\overline{U} = $ 0.04 (a) and 0.2 (b). Note that the quantities are presented in the $e_g$-orbital basis.}
    \label{sfig_gap_vertex_structure}
\end{figure}

\begin{figure} [!htbp] 
	\includegraphics[width=0.6\textwidth]{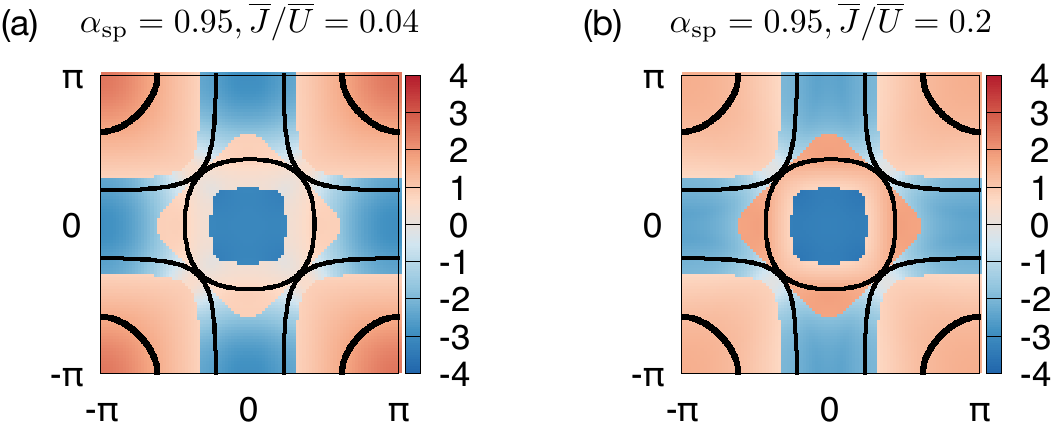}
	\caption{Gap functions $\hat{\Delta}_n(\bm{k},i\omega_0)$ (in meV) obtained by projecting $\Delta_{lm}(\bm{k},i\omega_0)$ at each $\bm{k}$ to the band $n$ closest to the Fermi level.  $\hat{\Delta}_n(\bm{k},i\omega_0)=\sum_{lm}P^{-1}_{nl}\Delta_{lm}(\bm{k},i\omega_0)P_{mn}$ where $P_{mn}$ ($P^{-1}_{nl}$ ) is the basis-transformation matrix from the $e_g$-orbital $m$ ($l$) to the band $n$ closest to the Fermi level. The black solid lines indicate the noninteracting FS obtained from DFT.}
	\label{sfig_gap_FS}
\end{figure}

Here, we discuss the full orbital and momentum structure of the dominant gap function $\Delta_{lm}(\bm{k},i\omega_0)$ and dominant matrix elements of the singlet pairing vertex $\Gamma^{\mathrm{s}}_{llmm}(\bm{q},i\nu_0)$ obtained from the CDMFT electronic structure. We show those in Fig.~\ref{sfig_gap_vertex_structure} for $\overline{J}/\overline{U} = 0.04$ and 0.2 at $\alpha_{\mathrm{sp}} = 0.95$ in the $e_g$-orbital basis; see Fig.~\ref{sfig_gap_FS} for the corresponding gap functions projected to the noninteracting FS, which shows clearly the $s \pm$-wave symmtery. The vertex generally has an orbital block structure with intralayer components being positive and interlayer components being negative which originates from the interlayer AFM fluctuations. The different dominant pairing vectors $\bm{Q}$ discussed in the main text and in Sec.~\ref{sec:Stoner_factors} can be easily distinguished by comparing panels (a) and (b). However, increasing $\overline{J}/\overline{U}$ does not only change the dominant $\bm{Q}$ vector, but it also changes the relative weight of the orbital components. Namely, for small $\overline{J}/\overline{U}$ mostly intraorbital components  $\Gamma^{\mathrm{s}}_{llll}$ and the interlayer $z$ components $\Gamma^{\mathrm{s}}_{\bar{z}\bar{z}\underline{z}\;\!\underline{z}}$ play a role, whereas for larger $\overline{J}/\overline{U}$ the magnitude of the components increases and evens out. This, in turn, affects the orbital structure of the dominant gap. By increasing $\overline{J}/\overline{U}$ , the gap function gains more weight in the interlayer $x$-component $\Delta_{\bar{x}\underline{x}}$, i.e., the gap opening on the FS pockets with $x_{\pm}$ character is enhanced.

\section{Additional data: The critical role of FM $\chi^0_{z_+ z_+  z_+ z_+}(q)$ in the singlet pairing}

\begin{figure*} [!htbp] 
	\includegraphics[width=1.0\textwidth]{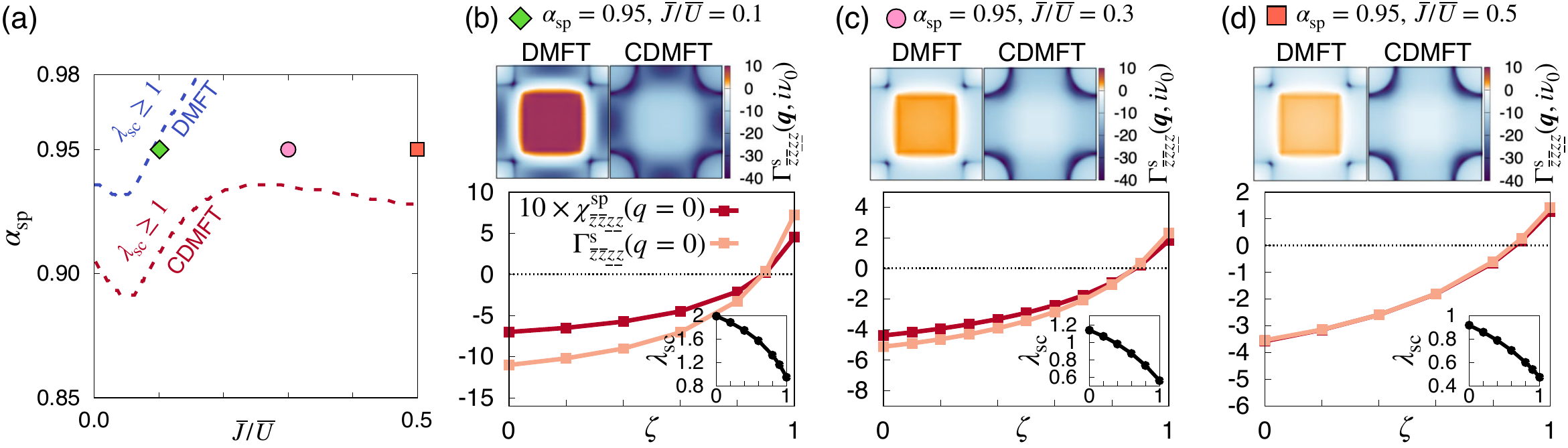}
	\caption{ (a) Superconducting phase diagram in the $\alpha_\mathrm{sp}$--$\overline{J}/ \overline{U}$ space at $T=1/145~\mathrm{eV} \simeq 80$~K.  The superconductivity sets in (i.e., $\lambda_\mathrm{sc} \geq 1$) in the regions above the dashed lines; blue for the DMFT and red for the CDMFT. (b--d) Upper panel: the spin-singlet pairing interaction $\Gamma^{\mathrm{s}}_{\bar{z} \bar{z} \underline{z} \underline{z}}(\bm{q},i\nu_0)$ between top and bottom layer $z$ orbitals in the FBZ. Lower panel: $\chi^{\mathrm{sp}}_{\bar{z} \bar{z} \underline{z} \underline{z}}(q=0)$, $\Gamma^{\mathrm{s}}_{\bar{z} \bar{z} \underline{z} \underline{z}}(q=0)$, and $\lambda_\mathrm{sc}$ (inset) as a function of scaling factor $\zeta$ for DMFT $\chi^0_{z_+ z_+  z_+ z_+}(q)$.  }
	\label{sfig_scaling}
\end{figure*}

As we have discussed in the main text, the disappearance of the $\gamma$ pocket from the FS due to IECs within CDMFT results leads to the suppression of the FM fluctuation arising from $\chi^0_{z_+ z_+  z_+ z_+}(\bm{q},i\nu_0)$. 
In the singlet channel, this FM fluctuation is directly manifested by a repulsive (rather than attractive) interaction $\Gamma^{\mathrm{s}}_{\bar{z} \bar{z} \underline{z} \underline{z}}(q=0)$. Quenching of the FM $\chi^0_{z_+ z_+  z_+ z_+}$ as in CDMFT yields the attractive pairing interaction $\Gamma^{\mathrm{s}}_{\bar{z} \bar{z} \underline{z} \underline{z}}(q=0)$ between top and bottom layer $z$ orbitals as clearly shown in Fig.~\ref{sfig_scaling}(b--d) for three distinct $\overline{J}/\overline{U}$ values. This, in turn, promotes the singlet pairing as evidenced by the enhanced superconducting instabilities in CDMFT [Fig.~\ref{sfig_scaling}(a)].

To further corroborate this argument, we analyze how the DMFT superconducting instabilities are affected by the FM fluctuation by introducing a scaling factor $\zeta$ for $\chi^0_{z_+ z_+  z_+ z_+}(q)$. Namely, $\chi^0_{z_+ z_+  z_+ z_+}(q)$ is rescaled to $\zeta \chi^0_{z_+ z_+  z_+ z_+}(q)$ before constructing $\bm{\chi}^{\mathrm{sp/ch}}$ and $\bm{\Gamma}^\mathrm{s}$, and then we monitor how $\lambda_\mathrm{sc}$ behaves due to this change. 

Interestingly, indeed, $\lambda_\mathrm{sc}$ increases with  decreasing $\zeta$ for all the $\overline{J}/\overline{U}$ values we investigated.
Looking into the related spin susceptibilities, the components involving solely $\bar{z}$ or $\underline{z}$ characters are found to be most affected by the rescaled $\chi^0_{z_+ z_+  z_+ z_+}(q)$ as expected. In effect, as presented in Fig.~\ref{sfig_scaling}(b--d), the interlayer FM spin susceptibility $\chi^\mathrm{sp}_{\bar{z} \bar{z} \underline{z} \underline{z}}(q=0)$ at $\zeta=1$ becomes AFM with decreasing $\zeta$. Through Eq.~(\ref{Gamma_s}) $\chi^\mathrm{sp}_{\bar{z} \bar{z} \underline{z} \underline{z}}(q)$ directly affects the corresponding pairing interaction $\Gamma^{\mathrm{s}}_{\bar{z} \bar{z} \underline{z} \underline{z}}(q)$, whereby it should faithfully follow the behavior of $\chi^\mathrm{sp}_{\bar{z} \bar{z} \underline{z}\underline{z}}(q)$. As such, the repulsive pairing interaction $\Gamma^{\mathrm{s}}_{\bar{z} \bar{z} \underline{z}\underline{z}}(q=0)$ at $\zeta=1$ turns attractive below $\zeta \simeq 0.9$ (for all the $\overline{J}/\overline{U}$ values) at which $\chi^\mathrm{sp}_{\bar{z} \bar{z} \underline{z} \underline{z}}(q=0)$ changes its sign; see orange lines in the lower panels of Fig.~\ref{sfig_scaling}(b--d).

\newpage
